\PassOptionsToPackage{numbers,compress}{natbib}
\documentclass{article}

\usepackage[preprint]{neurips_2025}

\usepackage[utf8]{inputenc}
\usepackage[T1]{fontenc}
\usepackage{hyperref}
\usepackage{url}
\usepackage{booktabs}
\usepackage{amsmath}
\usepackage{amsfonts}
\usepackage{amssymb}
\usepackage{graphicx}
\usepackage{microtype}

\title{What Medicine Taught Us About Fairness and What It Missed: Lessons from Reconsidering Race-Specific Lung Function Reference Algorithms}

\author{%
  Amin Adibi \\
  University of British Columbia \\
  Vancouver, BC, Canada \\
  \texttt{amin.adibi@ubc.ca} \\
  \And
  Mohsen Sadatsafavi \\
  University of British Columbia \\
  Vancouver, BC, Canada \\
  \texttt{mohsen.sadatsafavi@ubc.ca}
}

\begin{document}

\maketitle

\begin{abstract}
  Since 2019, medical societies have reconsidered race-specific clinical equations often in parallel to and largely independent from algorithmic fairness research. Focusing on lung function reference algorithms that affect medical care, insurance, and employment for hundreds of millions globally, we analyze the transition from race-specific GLI-2012 to race-averaged GLI-Global through a fairness lens. Drawing on historical context, citation analysis, and quantitative evaluation, we show (i) limited cross-citation between FAccT and clinical guideline revision efforts; (ii) that GLI-Global implicitly encodes assumptions about social determinants of health, behaving as if ~62\% of the Black--White gap in FEV\textsubscript{1} is exposure-related; and (iii) clinical validation studies operationalized a sufficiency-like fairness criterion long before its formalization in fairness literature, while neglecting foundational results such as the impossibility theorem has led to inefficiencies in clinical research. Overall, our analysis highlights the value of deeper, mutually beneficial engagement between medical and fairness communities and the public to accelerate progress toward equitable healthcare algorithms.

\end{abstract}

\section{Introduction}
At least 42 clinical calculators use race or ethnicity as predictors \cite{visweswaran_online_2025}. Since 2019, there has been a movement within medicine to reconsider the use of race in clinical equations and algorithms \cite{vyas_hidden_2020, nkinsi_how_2022}. The effort was initially led by medical student activists and gained momentum following the murder of George Floyd and the ensuing racial reckoning and resurgence of the \textit{Black Lives Matter} movement in the US \cite{powe_race_2024}. Over the years, various coalitions joined the movement \cite{khazanchi_nyc_2022, ross_regional_2024, council_of_medical_specialty_societies_together_2024} which convinced a number of professional medical societies to review their race-specific algorithms \cite{delgado_reassessing_2021, stephenson_task_2021, burnett-bowie_american_2024, wilson_role_2023}. As a result of these efforts, as of May 2026, clinical algorithms for lung function reference standards \cite{bhakta_race_2023, bowerman_race-neutral_2023}, kidney function \cite{delgado_unifying_2022}, cardiovascular risk for treating hypertension and dyslipidemia \cite{khan_development_2024, jones_2025, Morris_2026}, and vaginal birth after C-section \cite{the_american_college_of_obstetricians_and_gynecologists_practice_2021} have been revised by medical societies to remove race and ethnicity as explicit predictors. On the other hand, two professional medical societies have refused to remove race from their risk calculators used for thoracic surgery and osteoporotic fracture risk due to disproportionate accuracy loss that was deemed unfair to underserved populations \cite{macgillivray_impact_2023, shahian_social_2022,kanis_race-specific_2024}.

Although many studies have tried to tackle race in clinical algorithms in general, an emerging theme in reviews is that the issue is highly nuanced. Depending on the context and use case, algorithms can perpetuate or mitigate disparities regardless of whether they explicitly include race or ethnicity as input \cite{siddique_impact_2024, wilson_rethinking_2024}. Strategies to mitigate bias in algorithms can include both removal and addition of race or ethnicity as predictors \cite{cary_mitigating_2023}. One complicating factor is that whether an algorithm is aware of race is not always clear, particularly when unstructured or high-dimensional data are involved. Some AI models have been found to be able to predict race or ethnicity from chest X-rays or cardiac ultrasounds with high accuracy even though no human-readable race signal seems to exist in these images \cite{gichoya_ai_2022, zou_implications_2023, duffy_confounders_2022, adleberg_predicting_2022}.

Debates in medical literature around race in medical algorithms have happened in parallel and largely independent of the growing body of research on algorithmic fairness, accountability, and transparency (FAccT) in machine learning. Research in FAccT that was spearheaded by high profile cases such as suspected bias in the COMPAS recidivism score \cite{angwin_machine_2022, angwin_machine_2016} had a strong methodological focus and led to important insights, including the \textit{impossibility theorem} \cite{chouldechova_fair_2017, kleinberg_inherent_2016}. Previous FAccT contributions have engaged deeply with race and ethnicity in algorithms, including the issue of improper racial and ethnic categories \cite{mickel_racialethnic_2024, jaime_ethnic_2024} and proxy bias in healthcare \cite{obermeyer_dissecting_2019}. Many of these insights remain insufficiently integrated into clinical literature and vice versa due to insufficient back and forth between fairness and medical literature. As we show in Section~\ref{sec:disconnect}, only 1.32\% of citations in the FAccT papers of the last three years were to medical journals.

\paragraph{Contribution}
We analyze the transition from race-specific to race-averaged spirometry reference equations through the lens of algorithmic fairness. We (1) quantify limited cross-citation between recent FAccT proceedings and clinical medicine, and the absence of fairness-ML citations in key medical society reports; (2) formalize and quantify the implicit assumptions about social determinants of health (SDoH) embedded in race-averaged equations, finding it behaves as if ~62\% of the Black--White gap is due to SDoH in NHANES 2007--2012; (3) show how clinical outcome-validation studies operationalize a sufficiency-like criterion before its formalization in machine learning literature; (4) translate common spirometry scoring practices into independence/sufficiency terms and clarify the resulting trade-offs (including why race-neutral reference equations can yield race-aware scores); and (5) briefly review evidence on patient/public attitudes, highlighting transparency as a key design lever.

\section{The Disconnect Between Algorithmic Fairness and Medical Literature}
\label{sec:disconnect}
To quantify the engagement between the algorithmic fairness community and medical literature, we conducted a citation analysis of i) papers published in the ACM Conference on Fairness, Accountability, and Transparency (FAccT), and ii) taskforce reports from professional societies that have reconsidered the use of race and ethnicity in their clinical prediction models.

\subsection{Methods}

For the FAccT proceedings citation analysis, we included all archival papers from 2022 to 2024, inclusive. We obtained PDF papers from the ACM Digital Library proceedings for these three years, yielding 515 papers for analysis. We extracted bibliographic references from each paper using GROBID (v0.8.2), an open-source machine learning library for parsing scholarly documents \cite{GROBID}.  To identify citations to medical journals, we used the SCImago Journal \& Country Rank \cite{SCImago}, selecting the top 500 journals in the Medicine subject category ranked by the SCImago Journal Rank (SJR) indicator. General science journals (e.g., \textit{Nature}, \textit{Science}, \textit{PNAS}) were excluded to focus specifically on medical and clinical venues. 

To assess whether the disconnect is bidirectional, we examined citations in official medical society reports that recommended revising or removing race from clinical algorithms. We identified seven key policy documents: the American Thoracic Society (ATS) and European Respiratory Society (ERS) statement on race and ethnicity in pulmonary function testing \cite{bhakta_race_2023}, the National Kidney Foundation (NKF) and American Society of Nephrology (ASN) task force report on race in kidney function estimation \cite{delgado_unifying_2022, delgado_reassessing_2021}, the American Heart Association's scientific statement and its report on the development and validation of the PREVENT equations \cite{khan_development_2024, khan_novel_2023}, the American College of Obstetricians and Gynecologists (ACOG) practice advisory on vaginal birth after caesarean calculators \cite{the_american_college_of_obstetricians_and_gynecologists_practice_2021}, and the American Society for Bone and Mineral Research (ASBMR) task force report on fracture risk algorithms \cite{burnett-bowie_american_2024}.
For each document, we extracted all references and searched for citations to (1) papers published in ACM FAccT or its predecessor FAT* proceedings, (2) foundational algorithmic fairness papers on impossibility results \cite{chouldechova_fair_2017, kleinberg_inherent_2016}, FairML book \cite{barocas_fairness_2023}, seminal works on bias in healthcare algorithms \cite{obermeyer_dissecting_2019}, or any work from major machine learning venues such as NeurIPS, ICML, and AAAI. We additionally examined whether reports explicitly discussed fairness criteria (independence, separation, sufficiency, or equivalent concepts).

\subsection{Results}

Our analysis of FAccT proceedings included 39,203 total citations across 515 papers, representing 27,410 unique references. Of these citations, only 519 (1.32\%) referenced medical or clinical journals. When duplicates were removed, this represents 454 unique medical publications cited across the entire corpus. FAccT citations to medical literature were concentrated in a small number of venues (Table~\ref{tab:medical_venues}). \textit{Nature Medicine} was the most frequently cited medical journal (39 citations), followed by \textit{npj Digital Medicine} (29), \textit{The Lancet Digital Health} (28), and the \textit{Journal of the American Medical Informatics Association} (28). Only 47 medical papers were cited more than once across all FAccT proceedings analyzed, and none were cited more than 5 times.

\begin{table}[h]
\centering
\caption{Top 10 Medical Journals Cited in FAccT Proceedings (2022--2024)}
\label{tab:medical_venues}
\begin{tabular}{lr}
\toprule
\textbf{Journal} & \textbf{Citations} \\
\midrule
Nature Medicine & 39 \\
npj Digital Medicine & 29 \\
The Lancet Digital Health & 28 \\
Journal of the American Medical Informatics Association & 28 \\
New England Journal of Medicine & 19 \\
Journal of Medical Internet Research & 18 \\
American Journal of Public Health & 13 \\
JAMA Network Open & 13 \\
Journal of Biomedical Informatics & 12 \\
The Lancet & 11 \\
\bottomrule
\end{tabular}
\end{table}

None of the seven key policy documents by professional medical societies that discussed revisions to race-based algorithms had cited any FAccT papers or any papers in major machine learning conferences. Similarly, none used fairness terminology or cited any of the seminal works in fairness.

These findings underscore a substantial disconnect between the algorithmic fairness research community and clinical medicine. Despite shared concerns about bias, equity, and the deployment of predictive algorithms in healthcare settings, FAccT researchers cite medical journals at remarkably low rates and professional medical societies do not cite FAccT papers at all. Among medical papers cited in FAccT papers, the concentration of medical citations in digital health and medical informatics journals suggests that when engagement does occur, it tends to involve work already positioned at the technology-medicine interface, rather than the broader medical literature on health disparities, clinical decision-making, or patient outcomes.

\section{Spirometry}

Spirometry is one of the most common lung function tests performed to diagnose or monitor chronic lung and airways conditions. During the test, the individual takes a deep breath and then exhales forcefully into a mouthpiece that is connected to a machine called the spirometer. The machine records several metrics including forced expiratory volume in one second (FEV\textsubscript{1}), which is the volume of air expelled in the first second of exhalation, and forced vital capacity (FVC), the total volume of air exhaled. Unlike other physiological measures such as blood pressure, spirometric values are not immediately comparable between individuals; for instance, a taller person will naturally have bigger lungs and can exhale larger volumes, but that does not necessarily have any physiological implications. As such, spirometric values need to be normalized to be comparable. Measured volumes are normalized based on normative predicted values from reference algorithms. Spirometry reference algorithms estimate the expected healthy lung function values in an individual based on age, biological sex at birth, height, and sometimes race or ethnicity. Patients' measured spirometry values are compared to the \textit{predicted} values from reference algorithms to calculate a Z-score or divided by the predicted value to calculate percent predicted values. Guidelines typically suggest cut off thresholds based on the Z-score or percent predicted values along with consideration of clinical context for diagnosis, treatment planning, or severity ratings \cite{gold_2025, stanojevic_ersats_2022, bourbeau_2023_2023}.

\subsection{Early History}
Publications that discuss race and spirometry typically begin by acknowledging the racist motivations in 19th century applications of the device in the US, and connect that to the contemporary race-specific approaches to interpreting spirometry results. While spirometry has indeed been misused to advance a white supremacist agenda, this narrative misses nuanced differences in different applications of race adjustment, with vastly different fairness implications.

The spirometer emerged in Victorian England as a precision instrument for measuring vital capacity---the maximum volume of air that can be expelled from the lungs. While the history of the device goes back further, John Hutchinson's work in the 1840s established the spirometer as a credible scientific tool \cite{braun_breathing_2014}. In 1846, Hutchinson, who worked as a physician for a life insurance company, published results of spirometry tests on 2,130 individuals, categorized by their occupation, social status, and body types \cite{hutchinson_capacity_1846}. His enthusiastic promotion of the device's utility for life insurance assessments along the terms he coined (e.g. \textit{spirometer, vital capacity}) is a legacy that continues to this day \cite{bhakta_race_2023}.

\subsection{Scientific Racism and the Spirometer}

Discussions around racial differences in lung function go back centuries. In his only book, \textit{Notes on the State of Virginia}, Thomas Jefferson, the American Founding Father and third US president, claimed that Black and White individuals had structurally different lungs \cite{jefferson_notes_2022}. The explicit weaponization of spirometry to justify white supremacy happened a century later. Samuel Cartwright, a pro-slavery plantation physician was the first to report spirometry results by race and estimate what he deemed a ``deficiency'' of 20\% in the lungs of Black people \cite{braun_race_2023, braun_breathing_2014}. Following the American Civil War, Benjamin Gould's massive anthropometric study of Union soldiers devoted an entire chapter to lung capacity by race and reported 4-12\% lower average lung capacity for Black soldiers \cite{gould_investigations_1869}. Charles Darwin, in his widely read \textit{Descent of Man}, drew on Gould's data on lung capacity along with craniometry to argue for racial differences in humans \cite{darwin_descent_1981}.

Contemporary studies continue to show differences in average spirometry adjusted for sex, age, and height between different regions of the world \cite{duong_global_2013} and between groups of self-identified race and ethnicity in the US \cite{quanjer_multi-ethnic_2012}.

\subsection{Contemporary Race-Adjustment For Spirometry}

In 1974, Rossiter and Weill studied lung function in Black and White male asbestos workers and suggested that for clinical and epidemiological purposes that rely on normal values, a correction factor of 1.132 be multiplied by White reference standards to make them applicable to Black individuals \cite{rossiter_ethnic_1974}. This correction factor effectively treats lower lung function in Black individuals as benign, consistent with Rossiter and Weill's attribution of the differences to anthropometric differences.

The 1978 Occupational Safety and Health Administration (OSHA) Cotton Dust Standard mandated a race-specific scaling factor to \textit{prevent} employment discrimination against Black job applicants \cite{townsend_us_2022}. Current cotton dust standards recommend a race-specific reference equation \cite{occupational_safety_and_health_administration_29_2024}.

This history reveals parallel applications of race adjustment for clinical and occupational applications ---a duality that continues to complicate contemporary debates. As we show in later sections, adjusting for race in reference equations creates \textit{race unaware} metrics that preserve equal opportunity in employment screening, but may lead to underdiagnosis and undertreatment in clinical applications.

As of 1990, adjusting for race in reference values used in clinic was not the norm. A survey of 181 labs in Canada and the US found that 53\% of spirometry labs did not apply any ethnic correction. Of the remaining that did, 89\% applied a fixed correction factor ranging from 1.10 to 1.15 to different minority populations, while 11\% used population-specific references \cite{ghio_reference_1990}.

The Hankinson-1999 reference equations based on the Third National Health and Nutrition Examination Survey (NHANES III) were the first widely used race-specific references based on a representative sample of the US population \cite{hankinson_spirometric_1999}. In 2005, the American Thoracic Society (ATS) and the European Respiratory Society (ERS) started recommending race-specific reference equations  \cite{pellegrino_interpretative_2005}.

In 2012, the Global Lung Function Initiative (GLI) developed a set of \textit{multiethnic} spirometry reference equations to expand applicability of references beyond the US. GLI-2012 equations produced different predicted values for White, Black (African American), Northeast Asian, and Southeast Asian individuals \cite{quanjer_multi-ethnic_2012}. For individuals who did not identify with any of the four groups or were of mixed ethnic origin, a fifth category (GLI-Other) was created using the mathematical average of coefficients.

\subsection{Spirometry Reference Algorithms}

Reference algorithms affect whether a person's lung function will be interpreted as normal. The lower limit of normal (LLN) is conventionally defined as the 5th percentile of the reference population's distribution. This means that 5\% of apparently healthy individuals in the reference population will have values below the LLN and may be classified as \textit{abnormal} despite having no disease. Conversely, results close to predicted values do not necessarily exclude pulmonary pathology.

Critically, depending on the guideline and the jurisdiction, the choice of reference equations, along with the clinical observations may inform:
\begin{itemize}
    \item Whether a patient receives a diagnosis of obstructive or restrictive lung disease \cite{stanojevic_ersats_2022}
    \item The severity of chronic obstructive pulmonary disease (COPD) \cite{stanojevic_ersats_2022}
    \item Recommended pharmacotherapy for COPD \cite{bourbeau_2023_2023}
    \item Eligibility for occupations requiring respiratory fitness (e.g., firefighting, commercial driving) \cite{diao_implications_2024, Weissman2026, Baugh2026}
    \item Disability ratings and associated compensation payments \cite{khazanchi_reform_2025}
    \item Indication and priority for lung transplantation \cite{united_network_for_organ_sharing_guide_2020, shweish_indications_2019}
    \item Eligibility for surgical interventions such as lung cancer resection \cite{bonner_clinical_2023}
\end{itemize}
\subsection{From Race-Specific to Race-Averaged References}

In 2022, the race-averaged GLI-Global equations were developed using inverse probability weighting such that each of the four race/ethnicity groups contributes equally to the final equations, regardless of their representation in the original dataset \cite{bowerman_race-neutral_2023}. This approach weights observations from underrepresented groups (e.g., Black individuals) more heavily than an unweighted approach. The American Thoracic Society (ATS) and the European Respiratory Society (ERS) now recommend GLI-Global as the only reference equation for all patients regardless of race and ethnicity \cite{bhakta_race_2023}. This recommendation followed extensive debates about whether race-specific equations might underestimate the severity of lung disease in Black patients, perpetuate health disparities, and cause harm by delaying diagnoses and access to treatment.

\section{Critical Evaluation of Race-Averaged Spirometry References}

\subsection{Implicit Assumptions about Social Determinants of Health (SDoH)}

Race-specific spirometry reference equations normalize group differences in measured lung function. In clinical settings, this can be problematic: if lower average lung function reflects disproportionate harmful exposures (e.g., pollution, occupational risk, structural inequities), then treating those deficits as ``normal'' risks under-diagnosis and delayed access to care.

GLI-Global (2022) removes race as an explicit input and is often described as making no assumptions about the magnitude of SDoH effects \cite{bowerman_race-neutral_2023}. We argue, however, that any race-averaged reference implicitly encodes such assumptions through how it pools across groups. We formalize and estimate these implicit assumptions.

\subsubsection{A simple decomposition and an interpretable estimand}

We decompose measured lung function for the $i$th individual into an ``ideal'' component (in the absence of harmful exposures) minus a socio-environmental deficit:
\begin{equation}
\mathrm{LF}_i = \mathrm{LF}_i^{*} - D_i,
\end{equation}
where $D_i>0$ captures the cumulative impact of social and environmental exposures that prevent any individual from achieving their ideal lung function.

Let $E_i$ denote the race/ethnicity group of individual $i$, with $p$ a privileged reference group and $k$ another group of interest. Define the average observed gap:
\begin{equation}
\Delta_k = \mathbb{E}[\mathrm{LF}_i \mid E_i = p] - \mathbb{E}[\mathrm{LF}_i \mid E_i = k].
\end{equation}
Under the decomposition above, this gap can be conceptually viewed as a combination of differences in ideal physiology and differences in average exposure-related deficits. We summarize the latter with an interpretable parameter:
\begin{equation}
\phi_k \equiv \frac{\bar D_k - \bar D_p}{\Delta_k},
\end{equation}
i.e., the fraction of the observed gap attributable to differential SDoH-related deficit (with $\phi_k=1$ meaning ``entirely due to SDoH'' and $\phi_k=0$ meaning ``no average SDoH contribution'').

\paragraph{Implicit SDoH assumptions in a race-averaged reference.}
For an individual with covariates $X$, let $\widehat{\mathrm{LF}}^{[\mathrm{GLI\text{-}2012}]}_k(X)$ denote the group-$k$ predicted ``normal''
lung function from GLI-2012, and let $\widehat{\mathrm{LF}}^{[\mathrm{GLI\text{-}Global}]}_k(X)$ denote
the GLI-Global prediction for an individual in group $k$ (computed without using race).
For an individual with covariates $X$, we define the \emph{implicit SDoH assumption} encoded by GLI-Global
for group $k$ as:
\begin{equation}
\phi_k^{\mathrm{implicit}} \equiv
\frac{\widehat{\mathrm{LF}}^{[\mathrm{GLI\text{-}Global}]}_k(X) - \widehat{\mathrm{LF}}^{[\mathrm{GLI\text{-}2012}]}_k(X)}
{\widehat{\mathrm{LF}}^{[\mathrm{GLI\text{-}2012}]}_p(X) - \widehat{\mathrm{LF}}^{[\mathrm{GLI\text{-}2012}]}_k(X)}.
\end{equation}
This quantity measures how far GLI-Global moves from the group-$k$ prediction toward the privileged-group prediction, expressed as a fraction of the group-specific gap. If $\phi_k^{\mathrm{implicit}}$ matches the true $\phi_k$, then GLI-Global corresponds to the counterfactual where group $k$ experiences the privileged group's average SDoH-related deficit; deviations imply over- or under-correction.

\subsubsection{Methods}
We use NHANES 2007--2012 (adults 20--95 years of age with valid spirometry) and compute $\mathrm{FEV}_1$ z-scores under GLI-2012 (group-specific) and GLI-Global using the \texttt{rspiro} package.

To estimate $\phi_k^{\mathrm{implicit}}$, we construct a continuum of ``partially shifted'' predictions between the GLI-2012 group-$k$ and privileged-group-$p$ references by interpolating the predicted median $\widehat{\mathrm{LF}}$: $\widehat{\mathrm{LF}}_{\mathrm{adj}}(X;\phi)=\widehat{\mathrm{LF}}^{[\mathrm{GLI\text{-}2012}]}_k(X)+\phi(\widehat{\mathrm{LF}}^{[\mathrm{GLI\text{-}2012}]}_p(X)-\widehat{\mathrm{LF}}^{[\mathrm{GLI\text{-}2012}]}_k(X))$.
For each $\phi$, we compute corresponding $z$-scores using GLI-Global $(L,S)$ parameters and choose $\hat{\phi}_k^{\mathrm{implicit}}$ to minimize the mean squared error between the adjusted and GLI-Global $z$-scores within group $k$. Implementation details are in Appendix~\ref{app:phi_estimation}. We also run a sensitivity analysis using percent-predicted values.

\subsubsection{Results}

Our analysis included 9,878 adults: 4,183 non-Hispanic White, 2,089 non-Hispanic Black, 1,042 non-Hispanic Asian, and 2,564 Hispanic individuals. Non-Hispanic White and Hispanic individuals were mapped to the ``White'' category in GLI-2012, resulting in no racial gap to analyze; thus, our implicit SDoH estimation focused on non-Hispanic Black, non-Hispanic Asian, and Other Race (including multiracial) populations.

Table~\ref{tab:implicit_sdoh} presents the implicit SDoH assumptions embedded in GLI-Global equations by racial/ethnic group. For non-Hispanic Black adults, GLI-Global equations behaved as if approximately 62\% of the racial gap in lung function was attributable to SDoH factors ($\hat{\phi}_{\mathrm{Black}}^{\mathrm{implicit}} = 0.62$). For non-Hispanic Asian adults, the implicit SDoH assumption was approximately 39\% ($\hat{\phi}_{\mathrm{Asian}}^{\mathrm{implicit}} = 0.39$). For the ``Other Race'' category, which includes multiracial individuals, the implicit assumption was approximately 13\% ($\hat{\phi}_{\mathrm{Other}}^{\mathrm{implicit}} = 0.13$).

\begin{table}[h]
\centering
\caption{Implicit SDoH assumptions ($\hat{\phi}_k^{\mathrm{implicit}}$) in GLI-Global equations by race/ethnicity, expressed as percentages of the race-specific gap.}
\label{tab:implicit_sdoh}
\begin{tabular}{lc}
\toprule
Race/Ethnicity & $\hat{\phi}_k^{\mathrm{implicit}}$ (\%)\\
\midrule
Non-Hispanic Asian & 38.5 \\
Non-Hispanic Black & 62.1 \\
Other Race (Including Multiracial) & 12.5 \\
\bottomrule
\end{tabular}
\end{table}

\begin{figure}[h]
  \centering
  \includegraphics[width=\linewidth]{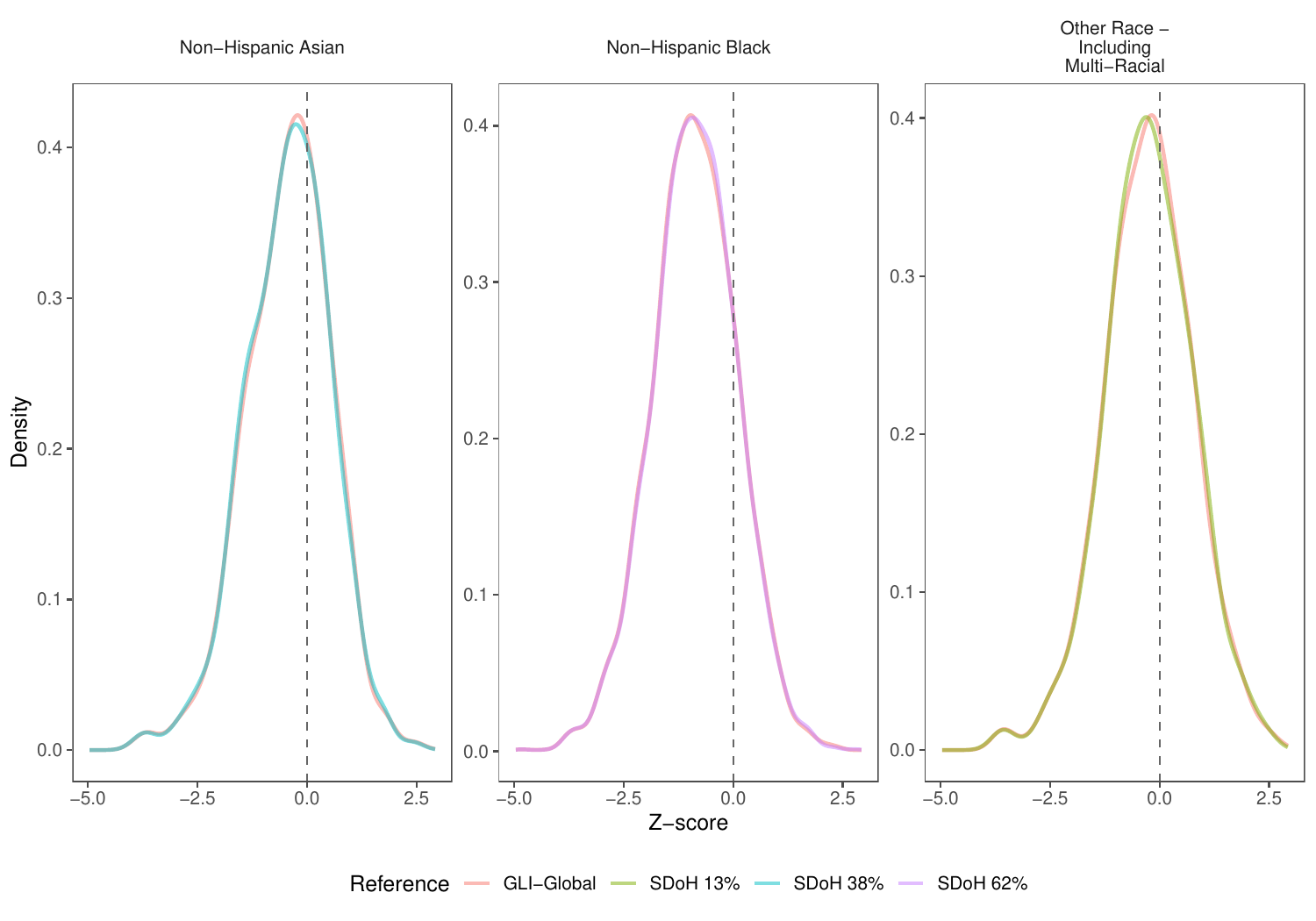}
  \caption{Comparison of GLI-Global FEV\textsubscript{1} Z-scores with Z-scores at the optimal SDoH percentage for minority racial/ethnic group in NHANES 2007-2012.}
  \label{fig:implicit_sdoh}
\end{figure}

Figure~\ref{fig:implicit_sdoh} displays the distribution of $z$-scores calculated under GLI-Global alongside $z$-scores calculated at the estimated $\hat{\phi}_k^{\mathrm{implicit}}$ for each group. The close overlap between distributions validates our calibration approach.

The sensitivity analysis using percent predicted $\mathrm{FEV}_1$ yielded similar results (Non-Hispanic Asian: 39.8\%, Non-Hispanic Black: 63.0\%, Other Race: 13.7\%).

\paragraph{Comparison to empirical estimates.}

These implicit assumptions can be compared to empirical estimates from the literature. Recent evidence suggests that 26.3\% of racial disparities in $\mathrm{FEV}_1$ between Black and White American adults and 6.6\% of the gap between Asian and White American adults are attributable to SDoH  ~\cite{adibi_social_2025}. GLI-Global's implicit assumption for Black ($\hat{\phi}_{\mathrm{Black}}^{\mathrm{implicit}} = 0.62$) and Asian adults ($\hat{\phi}_{\mathrm{Asian}}^{\mathrm{implicit}} = 0.39$) substantially exceeds these empirical estimates.

This discrepancy is not surprising: GLI-Global's inverse probability weighting methodology was designed to achieve equal representation across racial categories in the reference population, not to reflect empirically-determined contributions of social determinants to lung function differences. Our analysis makes explicit the SDoH assumptions that this methodological choice implicitly encodes.

\subsection{Prediction Models with No Truth Labels}
Validating reference equations is methodologically challenging. Spirometry reference algorithms are models that predict a counterfactual: healthy lung function in people who either have or are suspected of having lung diseases. By definition, these people do not have healthy lung function and as such, it is impossible to validate reference equations in the intended target population.

Historically, reference equations have been validated only internally or in other datasets consisting of \textit{healthy} volunteers who do not smoke and are not diagnosed with lung diseases such as asthma, COPD, and lung cancer \cite{quanjer_multi-ethnic_2012, hankinson_spirometric_1999}. However, this approach does not work for race-averaged or race-neutral reference algorithms, as the model will \textit{by design} be miscalibrated in subgroups of race or ethnicity \cite{bowerman_race-neutral_2023}.

Some recent studies have attempted to validate reference equations against clinical outcomes. Diao et al calculated concordance statistics (C statistics) for predicting respiratory symptoms, healthcare utilization, new-onset disease, and mortality using both GLI-2012 and GLI-Global equations \cite{diao_implications_2024}. Race-neutral and race-specific equations showed similar accuracy in predicting 11 clinical outcomes, indicating that removing race did not harm predictions but GLI-Global also fell short of expectations to improve outcome predictions. However, when reference equations exclude some variables due to fairness considerations (e.g., race or ethnicity), their association with outcomes might remain spuriously strong due to the confounding effect of such variables. As such, prediction accuracy \textit{per se} might become a misleading indicator of validity.

\subsubsection{Methods}
We analyzed data from the National Health and Nutrition Examination Survey (NHANES) III and NHANES 2007-2012, with mortality follow-up through December 31, 2019. We included participants with a history of smoking, diagnosis of a respiratory condition, or respiratory symptoms. We used area-under-the-curve (AUC) to compare discriminative accuracy in predicting multiple clinical outcomes using FEV\textsubscript{1} Z-scores from three alternative reference equations: race-averaged GLI-Global, race-specific GLI-2012, and a naïve equation that assigned the same predicted FEV\textsubscript{1} to all individuals. Evaluated clinical outcomes included 10-, 20-, and 30-year all-cause mortality, medical visits for wheezing, overnight hospital admissions within the past year, and concurrent respiratory symptoms.

\subsubsection{Results}
\begin{table*}[t]
\centering
\caption{Discriminative Accuracy of Reference-Adjusted FEV\textsubscript{1} Z-scores for Predicting Clinical Outcomes in Participants with or at Risk of Respiratory Diseases}
\label{tab:notruth}
\begin{tabular}{lccc}
\toprule
\textbf{Outcome and Cohort} & \multicolumn{3}{c}{\textbf{Discriminative Accuracy (95\% CI)*}} \\
\cmidrule(lr){2-4}
 & GLI-2012 & GLI-Global & Same FEV\textsubscript{1} for All\textsuperscript{\dag} \\
 & (Race-Specific) & (Race-Neutral) & (Na\"ive, Race-Neutral) \\
 & \multicolumn{3}{c}{Area-under-the-curve (AUC)} \\
\midrule
\textbf{Death, NHANES III} & & & \\
\quad 10-yr incidence from any cause & 0.604 (0.581, 0.626) & 0.640 (0.617, 0.662) & 0.787 (0.770, 0.804) \\
\quad 20-yr incidence from any cause & 0.610 (0.595, 0.625) & 0.642 (0.626, 0.658) & 0.788 (0.775, 0.800) \\
\quad 30-yr incidence from any cause & 0.568 (0.546, 0.590) & 0.592 (0.571, 0.613) & 0.747 (0.729, 0.764) \\
\addlinespace
\textbf{Recent healthcare utilization,} & & & \\
\textbf{NHANES 2007-2012} & & & \\
\quad Medical visit for wheezing in past yr & 0.525 (0.501, 0.549) & 0.520 (0.499, 0.541) & 0.565 (0.539, 0.591) \\
\quad Overnight hospital admission in past yr & 0.560 (0.537, 0.582) & 0.568 (0.544, 0.592) & 0.597 (0.576, 0.618) \\
\addlinespace
\textbf{Concurrent symptoms,} & & & \\
\textbf{NHANES 2007-2012} & & & \\
\quad Dyspnea on exertion & 0.643 (0.626, 0.660) & 0.649 (0.632, 0.666) & 0.653 (0.636, 0.670) \\
\quad Wheezing in past yr & 0.599 (0.584, 0.613) & 0.605 (0.591, 0.619) & 0.597 (0.582, 0.611) \\
\quad Cough for >3 months in past yr & 0.573 (0.549, 0.597) & 0.568 (0.545, 0.591) & 0.560 (0.537, 0.583) \\
\quad Dry cough at night in past yr & 0.515 (0.496, 0.534) & 0.522 (0.501, 0.542) & 0.562 (0.539, 0.584) \\
\quad Phlegm for >3 months in past yr & 0.591 (0.564, 0.618) & 0.592 (0.564, 0.619) & 0.583 (0.556, 0.610) \\
\bottomrule
\end{tabular}
\end{table*}
FEV\textsubscript{1} Z-scores from race-averaged GLI-Global were non-inferior to those from race-specific GLI-2012 in predicting clinical outcomes (Table~\ref{tab:notruth}). Similar results in the literature have been used to argue that race-averaged GLI-Global performs as well as GLI-2012 even after removing race as an explicit predictor. However, FEV\textsubscript{1} Z-scores based on the clearly flawed naïve reference equation showed significantly better accuracy in predicting death and comparable accuracy for healthcare utilization and symptoms among respiratory patients in NHANES, compared to GLI-2012 and GLI-Global (Table~\ref{tab:notruth}). Based on statistical assessment of prediction accuracy, one should advocate for using the naïve equation, which is clearly flawed.

\subsection{Race-Neutrality and the Impossibility Theorem}

We should note an important difference between spirometry reference algorithms and other models used in clinical practices. Unlike most settings in which a model's prediction is directly used for decision-making, spirometry reference models' output is first transformed into a Z-score or a percent predicted value and then used for decision-making.

In the context of spirometry:
\begin{itemize}
    \item The score $S$ could be a \textit{Z-score} or \textit{percent predicted} value (e.g., FEV\textsubscript{1} Z-scores) or a raw value (e.g., FVC)
    \item The sensitive attribute $A$ will be race or ethnicity
    \item Outcome $Y$ could be any clinical outcomes (e.g., symptoms, mortality, etc.)
\end{itemize}

The \textit{impossibility theorem} states that except in very special cases, fundamental fairness criteria are mathematically incompatible \cite{chouldechova_fair_2017, kleinberg_inherent_2016}. Let us now apply fundamental fairness criteria to spirometry metrics using race-specific and race-averaged reference equations. Let $Y$ be the outcome, $A$ the sensitive attribute (here race or ethnicity), and $S$ the score from the algorithm.

\paragraph{Independence ($S \perp A$)} This criterion is met when no information about the sensitive attribute is encoded in the score. Raw spirometry values such as FEV\textsubscript{1} and FVC correlate with race or ethnicity, and as such do not meet \textit{independence}. FEV\textsubscript{1} and FVC Z-scores and percent predicted values from race specific equations (e.g., GLI-2012 or NHANES-1999) do not correlate with race or ethnicity, and as such do meet \textit{independence} while those from race-averaged GLI-Global correlate with race and violate \textit{independence} \cite{adibi_impossibility_2025}. This is in contrast with other clinical prediction models such as eGFR or PREVENT, where the revised race-neutral version of the model meets \textit{independence}. Indeed, \textit{predicted} FEV\textsubscript{1} and FVC that are the immediate output of reference algorithms meet \textit{independence} if they are from revised race-averaged references, but it is the Z-score and percent predicted values, not the predicted reference values themselves, that are used for decision making.

\paragraph{Separation ($S \perp A \mid Y$)} In the context of classification, \textit{separation} is sometimes called \textit{error rate parity} as it requires equal false negative and false positive rates within the subgroups of the sensitive attribute. In the context of spirometry, this criterion translates to having the same distribution of race or ethnicity in groups who have experienced similar outcomes. Since \textit{separation} requires outcome, it can only be established after the fact. While it can be used for auditing purposes, its immediate clinical utility might be limited.

\paragraph{Sufficiency ($Y \perp A \mid S$)}
In the context of a continuous clinical prediction, a score that meets \textit{sufficiency} will lead to the same level of outcome, irrespective of the sensitive attribute. In the context of spirometry, this translates to the same score having the same prognostic implications for all patients regardless of race or ethnicity.

In 2012, Burney and Hooper showed that any given level of observed FVC raw is associated with the same mortality risk, regardless of whether the patient is Black or White \cite{burney_use_2012, burney_lung_2014}. Gaffney and colleagues repeated this study in the NHANES III population and found that the apparent higher mortality rate for Black participants disappeared after adjusting for either absolute FVC values or White-specific FVC percent predicted values \cite{gaffney_prognostic_2021}. More recently, Elmaleh-Sachs and colleagues showed that race-adjusted GLI 2012 equations for FEV\textsubscript{1} and FVC proved no more accurate than the race-average GLI 2012-Other in predicting hospitalizations and mortality attributed to lower respiratory disease \cite{elmaleh-sachs_raceethnicity_2022, schluger_vanishing_2022}. McCormack and colleagues have reported that GLI 2012-Other FEV\textsubscript{1} z-scores are a better predictor of long-term survival than race-specific FEV\textsubscript{1} z-scores in NHANES III \cite{mccormack_race_2022, bhakta_good_2022}. Ekström and Mannino showed that race-specific equations did not improve the prediction of breathlessness in NHANES 2007-2012 data \cite{ekstrom_race-specific_2022}. These studies form a large proportion of the evidence base that professional societies cited as justification for switching away from race-specific reference to race-averaged references \cite{bhakta_race_2023}.

What these studies collectively show is that raw spirometric values FEV\textsubscript{1} and FVC, along with their percent predicted or z-score variants from either race-averaged or single-race references meet the \textit{sufficiency} fairness criterion. In contrast, percent predicted or z-score variants from race-specific references violate \textit{sufficiency}. Interestingly, physicians and clinical researchers who conducted these studies had essentially discovered the \textit{sufficiency} fairness criterion, long before it was formalized by machine learning researchers \cite{barocas_fairness_2023}.

While the process to revise race-specific spirometry references was not guided by a formal fairness analysis, professional societies have for now recommended a solution that meets sufficiency, which appears appropriate for clinical care \cite{bhakta_race_2023}. However, spirometry has many use cases outside clinical applications, including for employment eligibility and insurance premiums.

Medical studies on implications of the race-averaged lung function references have shown, empirically, that removing race adjustment might lead to inequitable outcomes in some non-clinical applications \cite{diao_implications_2024}, an observation that would be mathematically obvious if one applies the impossibility theorem to spirometry metrics \cite{adibi_impossibility_2025}.

\subsection{Data Quality Issues}
The race-averaged GLI-Global reference algorithms are based on the same data as the older race-specific GLI-2012 and inherit the older dataset's quality issues \cite{bowerman_race-neutral_2023, quanjer_multi-ethnic_2012}. For one, the older dataset was based on the data collected for a variety of purposes and volunteered by different labs around the world, which makes the final dataset less reliable given the inherent variability of spirometry and the need for strict technical standardization \cite{stanojevic_ersats_2022}.

Another spillover issue from GLI-2012 is the handling of race labels and categories to calculate inverse probability weights in GLI-Global. The GLI-2012 dataset uses a combination of race labels from outdated self-identification and investigator assignment, with investigators assigning uniform racial categories to all participants from certain study sites \cite{quanjer_multi-ethnic_2012}. The race or ethnicity categories used in the study also seem to have been constructed based on human migration patterns and outdated theories of biological race, and ultimately combined into four categories of white, African American (with data only from the US), Southeast Asian and Northeast Asian, with the latter category later declared outdated by the investigators \cite{quanjer_secular_2015}. While creating an averaged equation with equal contribution from these four groups in the dataset does not seem to be supported by evidence \cite{wang_race_2024}, the investigators maintain that the weights have minimal impact \cite{bowerman_reply_2024}.

\subsection{Patient and Public Opinion}
A critical gap in the debate over race in clinical algorithms has been the relative absence of patient and public perspectives, a limitation acknowledged by the ATS \cite{bhakta_race_2023}. Recent studies  have begun to fill this gap. 

We recently conducted a multinational survey of 994 adults residing in 43 countries \cite{adibi_2026_survey} on public attitudes to algorithmic reform in medicine, with quota sampling to ensure approximately equal participation from ten groups of self-identified race and ethnicity. Most respondents (75.3\%) were comfortable with race or ethnicity as a predictor in clinical calculators, compared with only 47.1\% for postal or ZIP code; respondents were 2.5 times as likely to be uncomfortable with postal or ZIP code as with race or ethnicity, suggesting that socioeconomic proxies were not perceived as a more acceptable substitute. A large majority (87.9\%) were comfortable with race in an algorithm if their provider explained the rationale for using it, while only 12.8\% agreed that race and ethnicity should never be used in clinical decisions. For spirometry specifically, 40.0\% of respondents preferred the race-specific GLI-2012 equations, compared to 28.2\% preferred race-averaged GLI-Global equations. At the same time, a substantial proportion of respondents did not see themselves represented in the race and ethnicity categories provided in the existing race-specific algorithms (22.1\%--42.9\% depending on the algorithm) \cite{adibi_2026_survey}. 

A nationally-representative survey of 1,750 US adults by Diao et al.~\cite{diao_public_2025} also found similar patterns: 79.8\% of respondents preferred that their physician use all available information, including race, while only 20.1\% preferred that race be excluded. Another recent survey of 66 respirologists in the US found that only 9\% recommended the race-specific references compared to 64\% who preferred race-averaged spirometry reference equations \cite{brems2026}.

Collectively, these surveys, along with previous qualitative work by Schmidt et al.~\cite{schmidt_patients_2023} and Jain et al.~\cite{jain_awareness_2023} suggest that there might be a disconnect between professional consensus that increasingly favours race removal and public preferences largely accepting of race inclusion when transparent and clinically justified. Importantly, when patients are informed about the issues, their preferences appear more nuanced than a simple preference for or against race inclusion, with the adequacy of existing racial categories emerging as a distinct concern. It also appeared that substituting race with socioeconomic proxies such as area-based deprivation scores \cite{khan_development_2024} may not be perceived as more trustworthy by many patients.

\paragraph{Limitations}
Our analysis has several limitations. First, our quantitative analysis of implicit SDoH assumptions relies on NHANES 2007--2012 data, which may not be representative of current populations or other healthcare systems outside the US. The estimates of $\phi_k^{\mathrm{implicit}}$ should be interpreted as characterizations of GLI-Global's methodology rather than claims about true SDoH contributions to lung function differences.
Second, our analysis focuses on US racial categories, which do not translate directly to other contexts. The GLI equations are used globally, but their implications may differ substantially in countries with different racial demographics, healthcare systems, and histories of structural racism.
Third, we focus on spirometry as a case study; the lessons may not generalize to all clinical algorithms. Different algorithms have different purposes, different relationships between race and outcomes, and different stakeholder communities.

\section{Conclusion}
The spirometry case shows us that bridging the gap between medical and algorithmic fairness communities is a practical necessity. Physicians empirically discovered the sufficiency criterion years before its formalization in machine learning, yet impossibility results that could have clarified clinical debates remained obscure in medical literature. Our analysis demonstrates that the race-neutral references lead to race-aware Z-scores and encode quantifiable assumptions about social determinants of health. As other medical societies reconsider race in their algorithms, we hope the framework and findings presented here can help avoid repeating the missteps of parallel, siloed discourse.

\section*{Generative AI Usage Statement}
The authors have used generative AI (Claude Opus 4.5) to assist with formatting and grammar, and for debugging code.

\section*{Ethical Considerations Statement}
Participants provided informed consent per NHANES protocols \cite{centers_for_disease_control_and_prevention_cdc_national_nodate}. This study was exempt from the institutional Research Ethics Board review because it used only deidentified, publicly available data.

\newpage
\bibliographystyle{abbrvnat}
\bibliography{references}

@article{vyas_hidden_2020,
    title = {Hidden in {Plain} {Sight} — {Reconsidering} the {Use} of {Race} {Correction} in {Clinical} {Algorithms}},
    volume = {383},
    issn = {0028-4793},
    url = {https://doi.org/10.1056/NEJMms2004740},
    doi = {10.1056/NEJMms2004740},
    number = {9},
    urldate = {2023-11-07},
    journal = {New England Journal of Medicine},
    author = {Vyas, Darshali A. and Eisenstein, Leo G. and Jones, David S.},
    month = aug,
    year = {2020},
    note = {Publisher: Massachusetts Medical Society
\_eprint: https://doi.org/10.1056/NEJMms2004740},
    pages = {874--882},
}

@article{brems2026,
	title = {Current Practices and Preferences Regarding Race and Spirometry Interpretation},
	author = {Brems, J. Henry and Balasubramanian, Aparna and Ferryman, Kadija and Wharton, Robert and Eakin, Michelle N. and McCormack, Meredith C.},
	year = {2026},
	month = {02},
	date = {2026-02},
	journal = {CHEST Pulmonary},
	pages = {100239},
	doi = {10.1016/j.chpulm.2026.100239},
	url = {http://dx.doi.org/10.1016/j.chpulm.2026.100239},
	langid = {en}
}

@article{Weissman2026,
    author = {Weissman, David N and Mannino, David M and Laney, A Scott},
    title = {Spirometry for Assessing Work Ability and Impairment: Navigating Scylla and Charybdis},
    journal = {American Journal of Respiratory and Critical Care Medicine},
    pages = {aamag184},
    year = {2026},
    month = {04},
    issn = {1073-449X},
    doi = {10.1093/ajrccm/aamag184},
    url = {https://doi.org/10.1093/ajrccm/aamag184},
    eprint = {https://academic.oup.com/ajrccm/advance-article-pdf/doi/10.1093/ajrccm/aamag184/68020264/aamag184.pdf},
}

@article{Baugh2026,
    author = {Baugh, Aaron and Bhakta, Nirav R and Leduc, Todd, Chief(Ret) and Woodruff, Prescott G and Thakur, Neeta},
    title = {Understanding Firefighter Qualification Under Race-Neutral Spirometry \&amp; Evolving NFPA Guidelines},
    journal = {American Journal of Respiratory and Critical Care Medicine},
    pages = {aamag164},
    year = {2026},
    month = {04},
    issn = {1073-449X},
    doi = {10.1093/ajrccm/aamag164},
    url = {https://doi.org/10.1093/ajrccm/aamag164},
    eprint = {https://academic.oup.com/ajrccm/advance-article-pdf/doi/10.1093/ajrccm/aamag164/67782314/aamag164.pdf},
}

@article{Morris_2026,
	title = {2026 {ACC}/{AHA}/{AACVPR}/{ABC}/{ACPM}/{ADA}/{AGS}/{APhA}/{ASPC}/{NLA}/{PCNA} {Guideline} on the {Management} of {Dyslipidemia}: {A} {Report} of the {American} {College} of {Cardiology}/{American} {Heart} {Association} {Joint} {Committee} on {Clinical} {Practice} {Guidelines}},
	volume = {0},
	shorttitle = {2026 {ACC}/{AHA}/{AACVPR}/{ABC}/{ACPM}/{ADA}/{AGS}/{APhA}/{ASPC}/{NLA}/{PCNA} {Guideline} on the {Management} of {Dyslipidemia}},
	url = {https://www.ahajournals.org/doi/10.1161/CIR.0000000000001423},
	doi = {10.1161/CIR.0000000000001423},
	volume = {153},
	number = {17},
	year = {2026},
	urldate = {2026-04-16},
	journal = {Circulation},
	publisher = {American Heart Association},
	author = {{Writing Committee Members} and Blumenthal, Roger S. and Morris, Pamela B. and Gaudino, Mario and Johnson, Heather M. and Anderson, Timothy S. and Bittner, Vera A. and Blankstein, Ron and Brewer, LaPrincess C. and Cho, Leslie and de Ferranti, Sarah D. and Gianos, Eugenia and Gluckman, Ty J. and Gradney, Kristen F. and Isiadinso, Ijeoma and Lloyd-Jones, Donald M. and Marrs, Joel C. and Martin, Seth S. and McLain, Kellie H. and Mehta, Laxmi S. and Mora, Samia and Mulugeta, Wudeneh M. and Natarajan, Pradeep and Navar, Ann Marie and Orringer, Carl E. and Polonsky, Tamar S. and Reynolds, Harmony R. and Saseen, Joseph J. and Shapiro, Michael D. and Soffer, Daniel E. and Tynes, Sheila A. and Villavaso, Chloé D. and Virani, Salim S. and Wilkins, John T.},
}

@article {adibi_2026_survey,
	author = {Adibi, Amin and Le, Kristina Xingyu and Pierson, Emma and Diao, James A and Esfandiari, Nazafarin and Carlsten, Chris and Sadatsafavi, Mohsen},
	title = {Multinational Public Opinion on Race, Ethnicity, and Algorithmic Reform in Medicine},
	elocation-id = {2026.05.15.26352687},
	year = {2026},
	doi = {10.64898/2026.05.15.26352687},
	publisher = {Cold Spring Harbor Laboratory Press},
	URL = {https://www.medrxiv.org/content/early/2026/05/21/2026.05.15.26352687},
	eprint = {https://www.medrxiv.org/content/early/2026/05/21/2026.05.15.26352687.full.pdf},
	journal = {medRxiv}
}

@article{nkinsi_how_2022,
    title = {How the {University} of {Washington} {Implemented} a {Change} in {eGFR} {Reporting}},
    volume = {3},
    issn = {2641-7650},
    url = {https://www.ncbi.nlm.nih.gov/pmc/articles/PMC9034824/},
    doi = {10.34067/KID.0006522021},
    number = {3},
    urldate = {2024-01-22},
    journal = {Kidney360},
    author = {Nkinsi, Naomi T. and Young, Bessie A.},
    month = jan,
    year = {2022},
    pmid = {35582183},
    pmcid = {PMC9034824},
    pages = {557--560},
}

@article{powe_race_2024,
    title = {Race, {Health} {Care} {Algorithms}, and {Precision} {Health} {Equity}},
    issn = {0003-4819},
    url = {https://www.acpjournals.org/doi/full/10.7326/M24-0551},
    doi = {10.7326/M24-0551},
    urldate = {2024-03-14},
    journal = {Annals of Internal Medicine},
    author = {Powe, Neil R.},
    month = mar,
    year = {2024},
    note = {Publisher: American College of Physicians},
}

@article{bhakta_race_2023,
    title = {Race and {Ethnicity} in {Pulmonary} {Function} {Test} {Interpretation}: {An} {Official} {American} {Thoracic} {Society} {Statement}},
    volume = {207},
    issn = {1073-449X},
    shorttitle = {Race and {Ethnicity} in {Pulmonary} {Function} {Test} {Interpretation}},
    url = {https://www.atsjournals.org/doi/full/10.1164/rccm.202302-0310ST},
    doi = {10.1164/rccm.202302-0310ST},
    number = {8},
    urldate = {2023-11-10},
    journal = {American Journal of Respiratory and Critical Care Medicine},
    author = {Bhakta, Nirav R. and Bime, Christian and Kaminsky, David A. and McCormack, Meredith C. and Thakur, Neeta and Stanojevic, Sanja and Baugh, Aaron D. and Braun, Lundy and Lovinsky-Desir, Stephanie and Adamson, Rosemary and Witonsky, Jonathan and Wise, Robert A. and Levy, Sean D. and Brown, Robert and Forno, Erick and Cohen, Robyn T. and Johnson, Meshell and Balmes, John and Mageto, Yolanda and Lee, Cathryn T. and Masekela, Refiloe and Weiner, Daniel J. and Irvin, Charlie G. and Swenson, Erik R. and Rosenfeld, Margaret and Schwartzstein, Richard M. and Agrawal, Anurag and Neptune, Enid and Wisnivesky, Juan P. and Ortega, Victor E. and Burney, Peter},
    month = apr,
    year = {2023},
    pages = {978--995},
}

@article{delgado_unifying_2022,
    title = {A {Unifying} {Approach} for {GFR} {Estimation}: {Recommendations} of the {NKF}-{ASN} {Task} {Force} on {Reassessing} the {Inclusion} of {Race} in {Diagnosing} {Kidney} {Disease}},
    volume = {79},
    issn = {0272-6386, 1523-6838},
    shorttitle = {A {Unifying} {Approach} for {GFR} {Estimation}},
    url = {https://www.ajkd.org/article/S0272-6386(21)00828-3/fulltext},
    doi = {10.1053/j.ajkd.2021.08.003},
    language = {English},
    number = {2},
    urldate = {2023-12-18},
    journal = {American Journal of Kidney Diseases},
    author = {Delgado, Cynthia and Baweja, Mukta and Crews, Deidra C. and Eneanya, Nwamaka D. and Gadegbeku, Crystal A. and Inker, Lesley A. and Mendu, Mallika L. and Miller, W. Greg and Moxey-Mims, Marva M. and Roberts, Glenda V. and Peter, Wendy L. St and Warfield, Curtis and Powe, Neil R.},
    month = feb,
    year = {2022},
    pmid = {34563581},
    note = {Publisher: Elsevier},
    pages = {268--288.e1},
}

@article{khan_development_2024,
    title = {Development and {Validation} of the {American} {Heart} {Association}’s {PREVENT} {Equations}},
    volume = {149},
    url = {https://www.ahajournals.org/doi/10.1161/CIRCULATIONAHA.123.067626},
    doi = {10.1161/CIRCULATIONAHA.123.067626},
    number = {6},
    urldate = {2024-10-07},
    journal = {Circulation},
    author = {Khan, Sadiya S. and Matsushita, Kunihiro and Sang, Yingying and Ballew, Shoshana H. and Grams, Morgan E. and Surapaneni, Aditya and Blaha, Michael J. and Carson, April P. and Chang, Alexander R. and Ciemins, Elizabeth and Go, Alan S. and Gutierrez, Orlando M. and Hwang, Shih-Jen and Jassal, Simerjot K. and Kovesdy, Csaba P. and Lloyd-Jones, Donald M. and Shlipak, Michael G. and Palaniappan, Latha P. and Sperling, Laurence and Virani, Salim S. and Tuttle, Katherine and Neeland, Ian J. and Chow, Sheryl L. and Rangaswami, Janani and Pencina, Michael J. and Ndumele, Chiadi E. and Coresh, Josef and {for the Chronic Kidney Disease Prognosis Consortium and the American Heart Association Cardiovascular-Kidney-Metabolic Science Advisory Group}},
    month = feb,
    year = {2024},
    note = {Publisher: American Heart Association},
    pages = {430--449},
}

@misc{the_american_college_of_obstetricians_and_gynecologists_practice_2021,
    title = {Practice {Advisory}: {Counseling} {Regarding} {Approach} to {Delivery} {After} {Cesarean} and the {Use} of a {Vaginal} {Birth} {After} {Cesarean} {Calculator}},
    url = {https://www.acog.org/clinical/clinical-guidance/practice-advisory/articles/2021/12/counseling-regarding-approach-to-delivery-after-cesarean-and-the-use-of-a-vaginal-birth-after-cesarean-calculator},
    language = {en},
    urldate = {2025-07-03},
    author = {{the American College of Obstetricians and Gynecologists}},
    month = dec,
    year = {2021},
}

@article{siddique_impact_2024,
    title = {The {Impact} of {Health} {Care} {Algorithms} on {Racial} and {Ethnic} {Disparities}},
    volume = {177},
    issn = {0003-4819},
    url = {https://www.acpjournals.org/doi/10.7326/M23-2960},
    doi = {10.7326/M23-2960},
    number = {4},
    urldate = {2024-03-14},
    journal = {Annals of Internal Medicine},
    author = {Siddique, Shazia Mehmood and Tipton, Kelley and Leas, Brian and Jepson, Christopher and Aysola, Jaya and Cohen, Jordana B. and Flores, Emilia and Harhay, Michael O. and Schmidt, Harald and Weissman, Gary E. and Fricke, Julie and Treadwell, Jonathan R. and Mull, Nikhil K.},
    month = mar,
    year = {2024},
    note = {Publisher: American College of Physicians},
    pages = {484--496},
}

@book{wilson_rethinking_2024,
    address = {Washington, D.C.},
    title = {Rethinking {Race} and {Ethnicity} in {Biomedical} {Research}},
    isbn = {978-0-309-72463-0},
    url = {https://nap.nationalacademies.org/catalog/27913},
    language = {en},
    urldate = {2025-01-20},
    publisher = {National Academies Press},
    author = {{National Academies of Sciences, Engineering, and Medicine}},
    editor = {Wilson, M. Roy and Beachy, Sarah H. and Schumm, Samantha N.},
    year = {2024},
    doi = {10.17226/27913},
    note = {Pages: 27913},
}

@book{braun_breathing_2014,
    title = {Breathing {Race} into the {Machine}: {The} {Surprising} {Career} of the {Spirometer} from {Plantation} to {Genetics}},
    isbn = {978-1-4529-4100-4},
    shorttitle = {Breathing {Race} into the {Machine}},
    language = {en},
    publisher = {U of Minnesota Press},
    author = {Braun, Lundy},
    month = feb,
    year = {2014},
    note = {Google-Books-ID: My90DwAAQBAJ},
}

@article{townsend_us_2022,
    title = {U.{S}. {Occupational} {Historical} {Perspective} on {Race} and {Lung} {Function}},
    volume = {206},
    issn = {1073-449X},
    url = {https://www.atsjournals.org/doi/full/10.1164/rccm.202203-0565LE},
    doi = {10.1164/rccm.202203-0565LE},
    number = {6},
    urldate = {2024-01-07},
    journal = {American Journal of Respiratory and Critical Care Medicine},
    author = {Townsend, Mary C. and Cowl, Clayton T.},
    month = sep,
    year = {2022},
    note = {Publisher: American Thoracic Society - AJRCCM},
    pages = {789--790},
}

@article{diao_implications_2024,
    title = {Implications of {Race} {Adjustment} in {Lung}-{Function} {Equations}},
    volume = {390},
    url = {https://www.nejm.org/doi/full/10.1056/NEJMsa2311809},
    doi = {10.1056/NEJMsa2311809},
    number = {22},
    urldate = {2024-06-21},
    journal = {New England Journal of Medicine},
    author = {Diao, James A. and He, Yixuan and Khazanchi, R and Nguemeni Tiako, Max Jordan and {Witonsky Jonathan I.} and {Pierson Emma} and {Rajpurkar Pranav} and {Elhawary Jennifer R.} and {Melas-Kyriazi Luke} and {Yen Albert} and {Martin Alicia R.} and {Levy Sean} and {Patel Chirag J.} and {Farhat Maha} and {Borrell Luisa N.} and {Cho Michael H.} and {Silverman Edwin K.} and {Burchard Esteban G.} and {Manrai Arjun K.}},
    month = jun,
    year = {2024},
    note = {Publisher: Massachusetts Medical Society
\_eprint: https://www.nejm.org/doi/pdf/10.1056/NEJMsa2311809},
    pages = {2083--2097},
}

@article{stanojevic_ersats_2022,
    title = {{ERS}/{ATS} technical standard on interpretive strategies for routine lung function tests},
    volume = {60},
    copyright = {The content of this work is not subject to copyright. Design and branding are copyright ©ERS 2022. For reproduction rights and permissions contact permissions@ersnet.org. https://www.ersjournals.com/user-licence},
    issn = {0903-1936, 1399-3003},
    url = {https://erj.ersjournals.com/content/60/1/2101499},
    doi = {10.1183/13993003.01499-2021},
    language = {en},
    number = {1},
    urldate = {2023-12-05},
    journal = {European Respiratory Journal},
    author = {Stanojevic, Sanja and Kaminsky, David A. and Miller, Martin R. and Thompson, Bruce and Aliverti, Andrea and Barjaktarevic, Igor and Cooper, Brendan G. and Culver, Bruce and Derom, Eric and Hall, Graham L. and Hallstrand, Teal S. and Leuppi, Joerg D. and MacIntyre, Neil and McCormack, Meredith and Rosenfeld, Margaret and Swenson, Erik R.},
    month = jul,
    year = {2022},
    pmid = {34949706},
    note = {Publisher: European Respiratory Society
Section: ERS Official Documents},
}

@techreport{gold_2025,
    title = {Global {Strategy} for {Prevention}, diagnosis and {Management} of {COPD}: 2025 report},
    institution = {Global Strategy for the Diagnosis, Management, and Prevention of COPD},
    url = {https://goldcopd.org/2025-gold-report/},
    urldate = {2025-07-28},
    author = {{Global Initiative for Chronic Obstructive Lung Disease}},
    year = {2025},
}

@article{bourbeau_2023_2023,
    title = {2023 {Canadian} {Thoracic} {Society} {Guideline} on {Pharmacotherapy} in {Patients} with {Stable} {COPD}},
    volume = {7},
    issn = {2474-5332},
    url = {https://doi.org/10.1080/24745332.2023.2231451},
    doi = {10.1080/24745332.2023.2231451},
    number = {4},
    urldate = {2023-11-21},
    journal = {Canadian Journal of Respiratory, Critical Care, and Sleep Medicine},
    author = {Bourbeau, Jean and Bhutani, Mohit and Hernandez, Paul and Aaron, Shawn D. and Beauchesne, Marie-France and B. Kermelly, Sophie and D’Urzo, Anthony and Lal, Avtar and Maltais, François and Marciniuk, Jeffrey D. and Mulpuru, Sunita and Penz, Erika and Sin, Don D. and Van Dam, Anne and Wald, Joshua and Walker, Brandie L. and Marciniuk, Darcy D.},
    month = jul,
    year = {2023},
    note = {Publisher: Taylor \& Francis
\_eprint: https://doi.org/10.1080/24745332.2023.2231451},
    pages = {173--191},
}

@article{hankinson_spirometric_1999,
    title = {Spirometric {Reference} {Values} from a {Sample} of  the {General} {U}.{S}. {Population}},
    volume = {159},
    issn = {1073-449X},
    url = {https://www.atsjournals.org/doi/10.1164/ajrccm.159.1.9712108},
    doi = {10.1164/ajrccm.159.1.9712108},
    number = {1},
    urldate = {2022-11-07},
    journal = {American Journal of Respiratory and Critical Care Medicine},
    author = {Hankinson, John L. and Odencrantz, John R. and Fedan, Kathleen B.},
    month = jan,
    year = {1999},
    pages = {179--187},
}

@article{quanjer_multi-ethnic_2012,
    title = {Multi-ethnic reference values for spirometry for the 3–95-yr age range: the global lung function 2012 equations},
    volume = {40},
    copyright = {©ERS 2012},
    issn = {0903-1936, 1399-3003},
    shorttitle = {Multi-ethnic reference values for spirometry for the 3–95-yr age range},
    url = {https://erj.ersjournals.com/content/40/6/1324},
    doi = {10.1183/09031936.00080312},
    language = {en},
    number = {6},
    urldate = {2022-11-07},
    journal = {European Respiratory Journal},
    author = {Quanjer, Philip H. and Stanojevic, Sanja and Cole, Tim J. and Baur, Xaver and Hall, Graham L. and Culver, Bruce H. and Enright, Paul L. and Hankinson, John L. and Ip, Mary S. M. and Zheng, Jinping and Stocks, Janet and Initiative, the ERS Global Lung Function},
    month = dec,
    year = {2012},
    pmid = {22743675},
    pages = {1324--1343},
}

@article{bowerman_race-neutral_2023,
    title = {A {Race}-neutral {Approach} to the {Interpretation} of {Lung} {Function} {Measurements}},
    volume = {207},
    issn = {1073-449X},
    url = {https://www-atsjournals-org.eu1.proxy.openathens.net/doi/10.1164/rccm.202205-0963OC},
    doi = {10.1164/rccm.202205-0963OC},
    number = {6},
    urldate = {2023-11-06},
    journal = {American Journal of Respiratory and Critical Care Medicine},
    author = {Bowerman, Cole and Bhakta, Nirav R. and Brazzale, Danny and Cooper, Brendan R. and Cooper, Julie and Gochicoa-Rangel, Laura and Haynes, Jeffrey and Kaminsky, David A. and Lan, Le Thi Tuyet and Masekela, Refiloe and McCormack, Meredith C. and Steenbruggen, Irene and Stanojevic, Sanja},
    month = mar,
    year = {2023},
    note = {Publisher: American Thoracic Society - AJRCCM},
    pages = {768--774},
}

@inproceedings{jaime_ethnic_2024,
    address = {New York, NY, USA},
    series = {{FAccT} '24},
    title = {Ethnic {Classifications} in {Algorithmic} {Fairness}: {Concepts}, {Measures} and {Implications} in {Practice}},
    isbn = {979-8-4007-0450-5},
    shorttitle = {Ethnic {Classifications} in {Algorithmic} {Fairness}},
    url = {https://doi.org/10.1145/3630106.3658902},
    doi = {10.1145/3630106.3658902},
    urldate = {2026-01-04},
    booktitle = {Proceedings of the 2024 {ACM} {Conference} on {Fairness}, {Accountability}, and {Transparency}},
    publisher = {Association for Computing Machinery},
    author = {Jaime, Sofia and Kern, Christoph},
    month = jun,
    year = {2024},
    pages = {237--253},
}

@article{jones_2025,
    title = {2025 {AHA}/{ACC}/{AANP}/{AAPA}/{ABC}/{ACCP}/{ACPM}/{AGS}/{AMA}/{ASPC}/{NMA}/{PCNA}/{SGIM} {Guideline} for the {Prevention}, {Detection}, {Evaluation} and {Management} of {High} {Blood} {Pressure} in {Adults}: {A} {Report} of the {American} {College} of {Cardiology}/{American} {Heart} {Association} {Joint} {Committee} on {Clinical} {Practice} {Guidelines}},
    shorttitle = {2025 {AHA}/{ACC}/{AANP}/{AAPA}/{ABC}/{ACCP}/{ACPM}/{AGS}/{AMA}/{ASPC}/{NMA}/{PCNA}/{SGIM} {Guideline} for the {Prevention}, {Detection}, {Evaluation} and {Management} of {High} {Blood} {Pressure} in {Adults}},
    url = {https://www.ahajournals.org/doi/10.1161/HYP.0000000000000249},
    doi = {10.1161/HYP.0000000000000249},
    volume = {82},
	number = {10},
    urldate = {2025-08-28},
	year = {2025},
    journal = {Hypertension},
    author = {Jones, Daniel W. and Ferdinand, Keith C. and Taler, Sandra J. and Johnson, Heather M. and Shimbo, Daichi and Abdalla, Marwah and Altieri, M. Martine and Bansal, Nisha and Bello, Natalie A. and Bress, Adam P. and Carter, Jocelyn and Cohen, Jordana B. and Collins, Karen J. and Commodore-Mensah, Yvonne and Davis, Leslie L. and Egan, Brent and Khan, Sadiya S. and Lloyd-Jones, Donald M. and Melnyk, Bernadette Mazurek and Mistry, Eva A. and Ogunniyi, Modele O. and Schott, Stacey L. and Smith, Sidney C. and Talbot, Amy W. and Vongpatanasin, Wanpen and Watson, Karol E. and Whelton, Paul K. and Williamson, Jeff D.},
    note = {Publisher: American Heart Association},
}

@article{kanis_race-specific_2024,
    title = {Race-specific {FRAX} models are evidence-based and support equitable care: a response to the {ASBMR} {Task} {Force} report on {Clinical} {Algorithms} for {Fracture} {Risk}},
    volume = {35},
    issn = {1433-2965},
    shorttitle = {Race-specific {FRAX} models are evidence-based and support equitable care},
    url = {https://doi.org/10.1007/s00198-024-07162-w},
    doi = {10.1007/s00198-024-07162-w},
    language = {en},
    number = {9},
    urldate = {2026-01-05},
    journal = {Osteoporosis International},
    author = {Kanis, John A. and Harvey, Nicholas C. and Lorentzon, Mattias and Liu, Enwu and Schini, Marian and Abrahamsen, Bo and Adachi, Jonathan D. and Alokail, Majed and Borgstrom, Fredrik and Bruyère, Olivier and Carey, John J. and Clark, Patricia and Cooper, Cyrus and Curtis, Elizabeth M. and Dennison, Elaine M. and Díaz-Curiel, Manuel and Dimai, Hans P. and Grigorie, Daniel and Hiligsmann, Mickael and Khashayar, Patricia and Lems, Willem and Lewiecki, E. Michael and Lorenc, Roman S. and Papaioannou, Alexandra and Reginster, Jean-Yves and Rizzoli, René and Shiroma, Eric and Silverman, Stuart L. and Simonsick, Eleanor and Sosa-Henríquez, Manuel and Szulc, Pawel and Ward, Kate A. and Yoshimura, Noriko and Johansson, Helena and Vandenput, Liesbeth and McCloskey, Eugene V. and Gregson, Celia L. and Lau, Edith and Lips, Paul and Ortolani, Sergio and Papaioannou, Alexandra and Dawson-Hughes, Bess and Jiwa, Famida and on behalf of the Board of IOF, and the IOF Working Group on Epidemiology and Quality of Life},
    month = sep,
    year = {2024},
    pages = {1487--1496},
}

@article{burnett-bowie_american_2024,
    title = {The {American} {Society} for {Bone} and {Mineral} {Research} {Task} {Force} on clinical algorithms for fracture risk report},
    volume = {39},
    issn = {0884-0431},
    url = {https://doi.org/10.1093/jbmr/zjae048},
    doi = {10.1093/jbmr/zjae048},
    number = {5},
    urldate = {2026-01-05},
    journal = {Journal of Bone and Mineral Research},
    author = {Burnett-Bowie, Sherri-Ann M and Wright, Nicole C and Yu, Elaine W and Langsetmo, Lisa and Yearwood, Gabby M H and Crandall, Carolyn J and Leslie, William D and Cauley, Jane A},
    month = may,
    year = {2024},
    pages = {517--530},
}

@misc{khazanchi_nyc_2022,
    address = {Rochester, NY},
    type = {{SSRN} {Scholarly} {Paper}},
    title = {{NYC} {Coalition} to {End} {Racism} in {Clinical} {Algorithms} ({CERCA}) {Inaugural} {Report}},
    url = {https://papers.ssrn.com/abstract=4412122},
    language = {en},
    urldate = {2026-01-08},
    publisher = {Social Science Research Network},
    author = {Khazanchi, Rohan and Morse, Michelle},
    month = sep,
    year = {2022},
}

@article{ross_regional_2024,
    title = {The {Regional} {Coalition} to {Eliminate} {Race} {Based} {Medicine}},
    volume = {116},
    issn = {0027-9684},
    url = {https://www.sciencedirect.com/science/article/pii/S0027968424001469},
    doi = {10.1016/j.jnma.2024.07.065},
    number = {4},
    urldate = {2026-01-08},
    journal = {Journal of the National Medical Association},
    author = {Ross, Seun O. and Henry-Moss, Dare and Berkeley, Abiona and McBee, Dwight W.},
    month = aug,
    year = {2024},
    pages = {439},
}

@misc{council_of_medical_specialty_societies_together_2024,
    title = {Together to {Catalyze} {Change} for {Racial} {Equity} in {Clinical} {Algorithms}},
    url = {https://encodingequity.org/ttcc-key-learnings/},
    language = {en-US},
    urldate = {2026-01-07},
    journal = {Encoding Equity},
    author = {{Council of Medical Specialty Societies}},
    month = jun,
    year = {2024},
}

@article{delgado_reassessing_2021,
    title = {Reassessing the {Inclusion} of {Race} in {Diagnosing} {Kidney} {Diseases}: {An} {Interim} {Report} {From} the {NKF}-{ASN} {Task} {Force}.},
    volume = {78},
    copyright = {Copyright © 2021 National KidneyFoundation, Inc and the American Society of Nephrology. Published by Elsevier Inc. All rights reserved.},
    issn = {1523-6838 0272-6386},
    doi = {10.1053/j.ajkd.2021.03.008},
    language = {eng},
    number = {1},
    journal = {American journal of kidney diseases : the official journal of the National Kidney Foundation},
    author = {Delgado, Cynthia and Baweja, Mukta and Burrows, Nilka Ríos and Crews, Deidra C. and Eneanya, Nwamaka D. and Gadegbeku, Crystal A. and Inker, Lesley A. and Mendu, Mallika L. and Miller, W. Greg and Moxey-Mims, Marva M. and Roberts, Glenda V. and St Peter, Wendy L. and Warfield, Curtis and Powe, Neil R.},
    month = jul,
    year = {2021},
    pmid = {33845065},
    pmcid = {PMC8238889},
    note = {Place: United States},
    pages = {103--115},
}

@article{stephenson_task_2021,
    title = {Task {Force} {Advises} {Excluding} {Race} as {Factor} in {Estimates} of {Kidney} {Function}.},
    volume = {2},
    issn = {2689-0186},
    doi = {10.1001/jamahealthforum.2021.3788},
    language = {eng},
    number = {10},
    journal = {JAMA health forum},
    author = {Stephenson, Joan},
    month = oct,
    year = {2021},
    pmid = {36218889},
    note = {Place: United States},
    pages = {e213788},
}

@article{wilson_role_2023,
    title = {Role of {Race} in the {Interpretation} of {Pulmonary} {Function} {Tests}: {The} {American} {Thoracic} {Society}’s {Efforts} to {Mitigate} {Bias} in {Its} {Clinical} {Guidance}},
    volume = {207},
    issn = {1073-449X},
    shorttitle = {Role of {Race} in the {Interpretation} of {Pulmonary} {Function} {Tests}},
    url = {https://www.atsjournals.org/doi/10.1164/rccm.202303-0351ED},
    doi = {10.1164/rccm.202303-0351ED},
    number = {8},
    urldate = {2023-12-18},
    journal = {American Journal of Respiratory and Critical Care Medicine},
    author = {Wilson, Kevin C.},
    month = apr,
    year = {2023},
    note = {Publisher: American Thoracic Society - AJRCCM},
    pages = {961--962},
}

@misc{macgillivray_impact_2023,
    title = {Impact of {Healthcare} {Algorithms} on {Racial} and {Ethnic} {Disparities} in {Health} and {Healthcare}},
    url = {https://www.sts.org/sites/default/files/Advocacy/2023-03-08%20AHRQ%20Healthcare%20Algorithms%20Final.pdf},
    urldate = {2026-01-07},
    author = {MacGillivray, Thomas E},
    month = mar,
    year = {2023},
}

@article{shahian_social_2022,
    title = {Social {Risk} {Factors} in {Society} of {Thoracic} {Surgeons} {Risk} {Models}. {Part} 2: {Empirical} {Studies} in {Cardiac} {Surgery}; {Risk} {Model} {Recommendations}},
    volume = {113},
    issn = {1552-6259},
    shorttitle = {Social {Risk} {Factors} in {Society} of {Thoracic} {Surgeons} {Risk} {Models}. {Part} 2},
    doi = {10.1016/j.athoracsur.2021.11.069},
    language = {eng},
    number = {5},
    journal = {The Annals of Thoracic Surgery},
    author = {Shahian, David M. and Badhwar, Vinay and O'Brien, Sean M. and Habib, Robert H. and Han, Jane and McDonald, Donna E. and Antman, Mark S. and Higgins, Robert S. D. and Preventza, Ourania and Estrera, Anthony L. and Calhoon, John H. and Grondin, Sean C. and Cooke, David T.},
    month = may,
    year = {2022},
    pmid = {34998735},
    pages = {1718--1729},
}

@article{visweswaran_online_2025,
    title = {Online database of clinical algorithms with race and ethnicity},
    volume = {15},
    copyright = {2025 The Author(s)},
    issn = {2045-2322},
    url = {https://www.nature.com/articles/s41598-025-94152-5},
    doi = {10.1038/s41598-025-94152-5},
    language = {en},
    number = {1},
    urldate = {2026-01-04},
    journal = {Scientific Reports},
    author = {Visweswaran, Shyam and Sadhu, Eugene M. and Morris, Michele M. and Vis, Anushka R. and Samayamuthu, Malarkodi Jebathilagam},
    month = mar,
    year = {2025},
    note = {Publisher: Nature Publishing Group},
    pages = {10913},
}

@article{hutchinson_capacity_1846,
    title = {On the capacity of the lungs, and on the respiratory functions, with a view of establishing a precise and easy method of detecting disease by the spirometer},
    volume = {29},
    issn = {0959-5287},
    url = {https://pubmed.ncbi.nlm.nih.gov/20895846/},
    doi = {10.1177/095952874602900113},
    language = {en},
    urldate = {2025-03-17},
    journal = {Medico-chirurgical transactions},
    author = {Hutchinson, John},
    year = {1846},
    pmid = {20895846},
    note = {Publisher: Med Chir Trans},
}

@article{cary_mitigating_2023,
    title = {Mitigating {Racial} {And} {Ethnic} {Bias} {And} {Advancing} {Health} {Equity} {In} {Clinical} {Algorithms}: {A} {Scoping} {Review}},
    volume = {42},
    issn = {0278-2715},
    shorttitle = {Mitigating {Racial} {And} {Ethnic} {Bias} {And} {Advancing} {Health} {Equity} {In} {Clinical} {Algorithms}},
    url = {https://www.healthaffairs.org/doi/full/10.1377/hlthaff.2023.00553},
    doi = {10.1377/hlthaff.2023.00553},
    number = {10},
    urldate = {2024-01-19},
    journal = {Health Affairs},
    author = {Cary, Michael P. and Zink, Anna and Wei, Sijia and Olson, Andrew and Yan, Mengying and Senior, Rashaud and Bessias, Sophia and Gadhoumi, Kais and Jean-Pierre, Genevieve and Wang, Demy and Ledbetter, Leila S. and Economou-Zavlanos, Nicoleta J. and Obermeyer, Ziad and Pencina, Michael J.},
    month = oct,
    year = {2023},
    note = {Publisher: Health Affairs},
    pages = {1359--1368},
}

@article{gichoya_ai_2022,
    title = {{AI} recognition of patient race in medical imaging: a modelling study},
    volume = {4},
    issn = {2589-7500},
    shorttitle = {{AI} recognition of patient race in medical imaging},
    url = {https://www.thelancet.com/journals/landig/article/PIIS2589-7500(22)00063-2/fulltext},
    doi = {10.1016/S2589-7500(22)00063-2},
    language = {English},
    number = {6},
    urldate = {2023-06-24},
    journal = {The Lancet Digital Health},
    author = {Gichoya, Judy Wawira and Banerjee, Imon and Bhimireddy, Ananth Reddy and Burns, John L. and Celi, Leo Anthony and Chen, Li-Ching and Correa, Ramon and Dullerud, Natalie and Ghassemi, Marzyeh and Huang, Shih-Cheng and Kuo, Po-Chih and Lungren, Matthew P. and Palmer, Lyle J. and Price, Brandon J. and Purkayastha, Saptarshi and Pyrros, Ayis T. and Oakden-Rayner, Lauren and Okechukwu, Chima and Seyyed-Kalantari, Laleh and Trivedi, Hari and Wang, Ryan and Zaiman, Zachary and Zhang, Haoran},
    month = jun,
    year = {2022},
    pmid = {35568690},
    note = {Publisher: Elsevier},
    pages = {e406--e414},
}

@article{zou_implications_2023,
    title = {Implications of predicting race variables from medical images},
    volume = {381},
    url = {https://www.science.org/doi/full/10.1126/science.adh4260},
    doi = {10.1126/science.adh4260},
    number = {6654},
    urldate = {2023-07-18},
    journal = {Science},
    author = {Zou, James and Gichoya, Judy Wawira and Ho, Daniel E. and Obermeyer, Ziad},
    month = jul,
    year = {2023},
    note = {Publisher: American Association for the Advancement of Science},
    pages = {149--150},
}

@article{duffy_confounders_2022,
    title = {Confounders mediate {AI} prediction of demographics in medical imaging},
    volume = {5},
    copyright = {2022 The Author(s)},
    issn = {2398-6352},
    url = {https://www.nature.com/articles/s41746-022-00720-8},
    doi = {10.1038/s41746-022-00720-8},
    language = {en},
    number = {1},
    urldate = {2026-01-09},
    journal = {npj Digital Medicine},
    author = {Duffy, Grant and Clarke, Shoa L. and Christensen, Matthew and He, Bryan and Yuan, Neal and Cheng, Susan and Ouyang, David},
    month = dec,
    year = {2022},
    note = {Publisher: Nature Publishing Group},
    pages = {188},
}

@article{adleberg_predicting_2022,
    title = {Predicting {Patient} {Demographics} {From} {Chest} {Radiographs} {With} {Deep} {Learning}},
    volume = {19},
    issn = {1546-1440},
    url = {https://www.sciencedirect.com/science/article/pii/S1546144022005440},
    doi = {10.1016/j.jacr.2022.06.008},
    number = {10},
    urldate = {2026-01-09},
    journal = {Journal of the American College of Radiology},
    author = {Adleberg, Jason and Wardeh, Amr and Doo, Florence X. and Marinelli, Brett and Cook, Tessa S. and Mendelson, David S. and Kagen, Alexander},
    month = oct,
    year = {2022},
    pages = {1151--1161},
}

@incollection{angwin_machine_2022,
    title = {Machine {Bias}},
    booktitle = {Ethics of {Data} and {Analytics}},
    publisher = {Auerbach Publications},
    author = {Angwin, Julia and Larson, Jeff and Mattu, Surya and Kirchner, Lauren},
    year = {2022},
    note = {Num Pages: 11},
    pages = {254--264},
}

@article{angwin_machine_2016,
    title = {Machine {Bias}},
    url = {https://www.propublica.org/article/machine-bias-risk-assessments-in-criminal-sentencing},
    language = {en-US},
    urldate = {2026-01-09},
    journal = {ProPublica},
    author = {Angwin, Julia and Larson, Jeff and Mattu, Surya and Kirchner, Lauren},
    month = may,
    year = {2016},
}

@article{chouldechova_fair_2017,
    title = {Fair {Prediction} with {Disparate} {Impact}: {A} {Study} of {Bias} in {Recidivism} {Prediction} {Instruments}},
    volume = {5},
    issn = {2167-6461},
    shorttitle = {Fair {Prediction} with {Disparate} {Impact}},
    url = {https://www.liebertpub.com/doi/10.1089/big.2016.0047},
    doi = {10.1089/big.2016.0047},
    number = {2},
    urldate = {2023-12-19},
    journal = {Big Data},
    author = {Chouldechova, Alexandra},
    month = jun,
    year = {2017},
    note = {Publisher: Mary Ann Liebert, Inc., publishers},
    pages = {153--163},
}

@misc{kleinberg_inherent_2016,
    title = {Inherent {Trade}-{Offs} in the {Fair} {Determination} of {Risk} {Scores}},
    url = {http://arxiv.org/abs/1609.05807},
    doi = {10.48550/arXiv.1609.05807},
    urldate = {2023-12-19},
    publisher = {arXiv},
    author = {Kleinberg, Jon and Mullainathan, Sendhil and Raghavan, Manish},
    month = nov,
    year = {2016},
    note = {arXiv:1609.05807 [cs, stat]},
}

@book{gould_investigations_1869,
    address = {New York; Cambridge},
    title = {Investigations in the {Military} and {Anthropological} {Statistics} of {American} {Soldiers}},
    url = {https://link.gale.com/apps/doc/U0113046142/MOME?sid=summon&xid=50df57b0&pg=1},
    language = {English},
    urldate = {2023-11-21},
    publisher = {Published for the U. S. Sanitary Commission, by Hurd and Houghton; Riverside Press},
    author = {Gould, Benjamin Apthorp},
    year = {1869},
}

@article{khazanchi_reform_2025,
    title = {Reform and {Remedy} for {Imprecision} and {Inequity} — {Ending} the {Race}-{Based} {Evaluation} of {Occupational} {Pulmonary} {Impairment}},
    volume = {393},
    issn = {0028-4793},
    url = {https://www.nejm.org/doi/full/10.1056/NEJMms2416661},
    doi = {10.1056/NEJMms2416661},
    number = {5},
    urldate = {2025-08-02},
    journal = {New England Journal of Medicine},
    author = {Khazanchi, Rohan and Stanojevic, Sanja and Hines, Stella E. and Bhakta, Nirav R.},
    month = jul,
    year = {2025},
    note = {Publisher: Massachusetts Medical Society
\_eprint: https://www.nejm.org/doi/pdf/10.1056/NEJMms2416661},
    pages = {508--514},
}

@techreport{united_network_for_organ_sharing_guide_2020,
    title = {A guide to calculating the lung allocation score},
    url = {https://unos.org/wp-content/uploads/unos/lung-allocation-score.pdf},
    language = {en},
    urldate = {2026-01-09},
    author = {{United Network for Organ Sharing}},
    month = jul,
    year = {2020},
}

@article{bonner_clinical_2023,
    title = {Clinical {Implications} of {Removing} {Race}-{Corrected} {Pulmonary} {Function} {Tests} for {African} {American} {Patients} {Requiring} {Surgery} for {Lung} {Cancer}},
    volume = {158},
    issn = {2168-6254},
    url = {https://doi.org/10.1001/jamasurg.2023.3239},
    doi = {10.1001/jamasurg.2023.3239},
    number = {10},
    urldate = {2024-01-21},
    journal = {JAMA Surgery},
    author = {Bonner, Sidra N. and Lagisetty, Kiran and Reddy, Rishindra M. and Engeda, Yadonay and Griggs, Jennifer J. and Valley, Thomas S.},
    month = oct,
    year = {2023},
    pages = {1061--1068},
}

@article{shweish_indications_2019,
    title = {Indications for lung transplant referral and listing},
    volume = {11},
    issn = {2072-1439},
    doi = {10.21037/jtd.2019.05.09},
    language = {eng},
    number = {Suppl 14},
    journal = {Journal of Thoracic Disease},
    author = {Shweish, Omar and Dronavalli, Goutham},
    month = sep,
    year = {2019},
    pmid = {31632748},
    pmcid = {PMC6783714},
    pages = {S1708--S1720},
}

@article{burney_use_2012,
    title = {The use of ethnically specific norms for ventilatory function in {African}-{American} and white populations},
    volume = {41},
    issn = {0300-5771},
    url = {https://doi.org/10.1093/ije/dys011},
    doi = {10.1093/ije/dys011},
    number = {3},
    urldate = {2023-11-16},
    journal = {International Journal of Epidemiology},
    author = {Burney, PGJ and Hooper, RL},
    month = jun,
    year = {2012},
    pages = {782--790},
}

@article{burney_lung_2014,
    title = {Lung function, genetics and ethnicity},
    volume = {43},
    issn = {0903-1936, 1399-3003},
    url = {https://publications.ersnet.org/content/erj/43/2/340},
    doi = {10.1183/09031936.00179313},
    language = {EN},
    number = {2},
    urldate = {2025-09-16},
    journal = {European Respiratory Journal},
    author = {Burney, Peter and Hooper, Richard},
    month = jan,
    year = {2014},
    note = {Publisher: European Respiratory Society
Section: Editorials},
    pages = {340--342},
}

@article{gaffney_prognostic_2021,
    title = {Prognostic implications of differences in forced vital capacity in black and white {US} adults: {Findings} from {NHANES} {III} with long-term mortality follow-up},
    volume = {39},
    issn = {2589-5370},
    shorttitle = {Prognostic implications of differences in forced vital capacity in black and white {US} adults},
    url = {https://www.thelancet.com/journals/eclinm/article/PIIS2589-5370(21)00353-9/fulltext},
    doi = {10.1016/j.eclinm.2021.101073},
    language = {English},
    urldate = {2024-06-15},
    journal = {eClinicalMedicine},
    author = {Gaffney, Adam W. and McCormick, Danny and Woolhandler, Steffie and Christiani, David C. and Himmelstein, David U.},
    month = sep,
    year = {2021},
    pmid = {34458707},
    note = {Publisher: Elsevier},
}

@article{elmaleh-sachs_raceethnicity_2022,
    title = {Race/{Ethnicity}, {Spirometry} {Reference} {Equations}, and {Prediction} of {Incident} {Clinical} {Events}: {The} {Multi}-{Ethnic} {Study} of {Atherosclerosis} ({MESA}) {Lung} {Study}},
    volume = {205},
    issn = {1073-449X},
    shorttitle = {Race/{Ethnicity}, {Spirometry} {Reference} {Equations}, and {Prediction} of {Incident} {Clinical} {Events}},
    url = {https://www.atsjournals.org/doi/10.1164/rccm.202107-1612OC},
    doi = {10.1164/rccm.202107-1612OC},
    number = {6},
    urldate = {2023-11-22},
    journal = {American Journal of Respiratory and Critical Care Medicine},
    author = {Elmaleh-Sachs, Arielle and Balte, Pallavi and Oelsner, Elizabeth C. and Allen, Norrina B. and Baugh, Aaron and Bertoni, Alain G. and Hankinson, John L. and Pankow, Jim and Post, Wendy S. and Schwartz, Joseph E. and Smith, Benjamin M. and Watson, Karol and Barr, R. Graham},
    month = mar,
    year = {2022},
    note = {Publisher: American Thoracic Society - AJRCCM},
    pages = {700--710},
}

@article{schluger_vanishing_2022,
    title = {The {Vanishing} {Rationale} for the {Race} {Adjustment} in {Pulmonary} {Function} {Test} {Interpretation}},
    volume = {205},
    issn = {1073-449X},
    url = {https://www.atsjournals.org/doi/10.1164/rccm.202112-2772ED},
    doi = {10.1164/rccm.202112-2772ED},
    number = {6},
    urldate = {2023-11-22},
    journal = {American Journal of Respiratory and Critical Care Medicine},
    author = {Schluger, Neil W.},
    month = mar,
    year = {2022},
    note = {Publisher: American Thoracic Society - AJRCCM},
    pages = {612--614},
}

@article{mccormack_race_2022,
    title = {Race, {Lung} {Function}, and {Long}-{Term} {Mortality} in the {National} {Health} and {Nutrition} {Examination} {Survey} {III}},
    volume = {205},
    issn = {1073-449X},
    url = {https://www.atsjournals.org/doi/full/10.1164/rccm.202104-0822LE},
    doi = {10.1164/rccm.202104-0822LE},
    number = {6},
    urldate = {2023-12-05},
    journal = {American Journal of Respiratory and Critical Care Medicine},
    author = {McCormack, Meredith C. and Balasubramanian, Aparna and Matsui, Elizabeth C. and Peng, Roger D. and Wise, Robert A. and Keet, Corinne A.},
    month = mar,
    year = {2022},
    note = {Publisher: American Thoracic Society - AJRCCM},
    pages = {723--724},
}

@article{bhakta_good_2022,
    title = {A {Good} {Fit} versus {One} {Size} for {All}: {Alternatives} to {Race} in the {Interpretation} of {Pulmonary} {Function} {Tests}},
    volume = {205},
    issn = {1073-449X},
    shorttitle = {A {Good} {Fit} versus {One} {Size} for {All}},
    url = {https://www.atsjournals.org/doi/10.1164/rccm.202201-0076ED},
    doi = {10.1164/rccm.202201-0076ED},
    number = {6},
    urldate = {2023-12-31},
    journal = {American Journal of Respiratory and Critical Care Medicine},
    author = {Bhakta, Nirav R. and Balmes, John R.},
    month = mar,
    year = {2022},
    note = {Publisher: American Thoracic Society - AJRCCM},
    pages = {616--618},
}

@article{ekstrom_race-specific_2022,
    title = {Race-specific reference values and lung function impairment, breathlessness and prognosis: {Analysis} of {NHANES} 2007–2012},
    volume = {23},
    issn = {1465-993X},
    shorttitle = {Race-specific reference values and lung function impairment, breathlessness and prognosis},
    url = {https://doi.org/10.1186/s12931-022-02194-4},
    doi = {10.1186/s12931-022-02194-4},
    number = {1},
    urldate = {2023-12-19},
    journal = {Respiratory Research},
    author = {Ekström, Magnus and Mannino, David},
    month = oct,
    year = {2022},
    pages = {271},
}

@article{adibi_social_2025,
    title = {Social {Determinants} of {Health} and {Racial} {Disparities} in {Lung} {Function}: {Findings} from {NHANES} 2007-2012},
    issn = {1535-4970},
    shorttitle = {Social {Determinants} of {Health} and {Racial} {Disparities} in {Lung} {Function}},
    doi = {10.1164/rccm.202501-0280OC},
    language = {eng},
    journal = {American Journal of Respiratory and Critical Care Medicine},
    author = {Adibi, Amin and Carlsten, Christopher and Brigham, Emily P. and Sin, Don D. and Loewen, Peter and Sadatsafavi, Mohsen},
    month = may,
    year = {2025},
    pmid = {40343923},
}

@book{barocas_fairness_2023,
    title = {Fairness and {Machine} {Learning}: {Limitations} and {Opportunities}},
    isbn = {978-0-262-37652-5},
    shorttitle = {Fairness and {Machine} {Learning}},
    language = {en},
    publisher = {MIT Press},
    author = {Barocas, Solon and Hardt, Moritz and Narayanan, Arvind},
    month = dec,
    year = {2023},
    note = {Google-Books-ID: HuGwEAAAQBAJ},
}

@article{adibi_impossibility_2025,
    title = {Is {Achieving} a {Fully} {Race}-{Neutral} {Approach} to {Lung} {Function} {Classification} {Even} {Possible}?},
    volume = {211},
    issn = {1073-449X},
    url = {https://www.atsjournals.org/doi/full/10.1164/rccm.202408-1599VP},
    doi = {10.1164/rccm.202408-1599VP},
    number = {3},
    urldate = {2025-06-26},
    journal = {American Journal of Respiratory and Critical Care Medicine},
    author = {Adibi, Amin and Sadatsafavi, Mohsen and Brigham, Emily P. and Bhatt, Surya P.},
    month = mar,
    year = {2025},
    note = {Publisher: American Thoracic Society - AJRCCM},
    pages = {432--435},
}

@article{wang_race_2024,
    title = {The {Race} {Arithmetic} of the {Global} {Lung} {Function} {Initiative} {Global} {Reference} {Equations}},
    volume = {209},
    issn = {1073-449X},
    url = {https://www.atsjournals.org/doi/10.1164/rccm.202303-0565LE},
    doi = {10.1164/rccm.202303-0565LE},
    number = {1},
    urldate = {2024-01-21},
    journal = {American Journal of Respiratory and Critical Care Medicine},
    author = {Wang, Richard J.},
    month = jan,
    year = {2024},
    note = {Publisher: American Thoracic Society - AJRCCM},
    pages = {112--113},
}

@article{quanjer_secular_2015,
    title = {Secular {Changes} in {Relative} {Leg} {Length} {Confound} {Height}-{Based} {Spirometric} {Reference} {Values}},
    volume = {147},
    issn = {0012-3692},
    url = {https://journal.chestnet.org/article/S0012-3692(15)39677-X/abstract},
    doi = {10.1378/chest.14-1365},
    language = {English},
    number = {3},
    urldate = {2025-03-25},
    journal = {CHEST},
    author = {Quanjer, Philip H. and Kubota, Masaru and Kobayashi, Hirosuke and Omori, Hisamitsu and Tatsumi, Koichiro and Kanazawa, Minoru and Stanojevic, Sanja and Stocks, Janet and Cole, Tim J.},
    month = mar,
    year = {2015},
    pmid = {25254426},
    note = {Publisher: Elsevier},
    pages = {792--797},
}

@article{bowerman_reply_2024,
    title = {Reply to: {The} {Race} {Arithmetic} of the {Global} {Lung} {Function} {Initiative} {Global} {Reference} {Equations}},
    volume = {209},
    issn = {1073-449X},
    shorttitle = {Reply to},
    url = {https://www.atsjournals.org/doi/10.1164/rccm.202304-0729LE},
    doi = {10.1164/rccm.202304-0729LE},
    number = {1},
    urldate = {2024-01-21},
    journal = {American Journal of Respiratory and Critical Care Medicine},
    author = {Bowerman, Cole and Bhakta, Nirav R. and Brazzale, Danny and Cooper, Brendan G. and Cooper, Julie and Gochicoa-Rangel, Laura and Haynes, Jeffrey and Kaminsky, David A. and Lan, Le Thi Tuyet and Masekela, Refiloe and McCormack, Meredith C. and Steenbruggen, Irene and Stanojevic, Sanja},
    month = jan,
    year = {2024},
    note = {Publisher: American Thoracic Society - AJRCCM},
    pages = {114--115},
}

@inproceedings{mickel_racialethnic_2024,
    address = {New York, NY, USA},
    series = {{FAccT} '24},
    title = {Racial/{Ethnic} {Categories} in {AI} and {Algorithmic} {Fairness}: {Why} {They} {Matter} and {What} {They} {Represent}},
    isbn = {979-8-4007-0450-5},
    shorttitle = {Racial/{Ethnic} {Categories} in {AI} and {Algorithmic} {Fairness}},
    url = {https://doi.org/10.1145/3630106.3659050},
    doi = {10.1145/3630106.3659050},
    urldate = {2026-01-04},
    booktitle = {Proceedings of the 2024 {ACM} {Conference} on {Fairness}, {Accountability}, and {Transparency}},
    publisher = {Association for Computing Machinery},
    author = {Mickel, Jennifer},
    month = jun,
    year = {2024},
    pages = {2484--2494},
}

@article{obermeyer_dissecting_2019,
    title = {Dissecting racial bias in an algorithm used to manage the health of populations},
    volume = {366},
    issn = {1095-9203},
    doi = {10.1126/science.aax2342},
    language = {eng},
    number = {6464},
    journal = {Science},
    author = {Obermeyer, Ziad and Powers, Brian and Vogeli, Christine and Mullainathan, Sendhil},
    month = oct,
    year = {2019},
    pmid = {31649194},
    pages = {447--453},
}

@article{schmidt_patients_2023,
    title = {Patients’ {Perspectives} on {Race} and the {Use} of {Race}-{Based} {Algorithms} in {Clinical} {Decision}-{Making}: a {Qualitative} {Study}},
    volume = {38},
    issn = {1525-1497},
    shorttitle = {Patients’ {Perspectives} on {Race} and the {Use} of {Race}-{Based} {Algorithms} in {Clinical} {Decision}-{Making}},
    url = {https://doi.org/10.1007/s11606-023-08035-4},
    doi = {10.1007/s11606-023-08035-4},
    language = {en},
    number = {9},
    urldate = {2024-04-05},
    journal = {Journal of General Internal Medicine},
    author = {Schmidt, Insa M. and Shohet, Merav and Serrano, Mariana and Yadati, Pranav and Menn-Josephy, Hanni and Ilori, Titilayo and Eneanya, Nwamaka D. and Cleveland Manchanda, Emily C. and Waikar, Sushrut S.},
    month = jul,
    year = {2023},
    pages = {2045--2051},
}

@misc{GROBID,
    title = {GROBID},
    author = {Lopez, Patrice},
    howpublished = {\url{https://github.com/kermitt2/grobid}},
    publisher = {GitHub},
    year = {2008--2025},
    archivePrefix = {swh},
    eprint = {1:dir:dab86b296e3c3216e2241968f0d63b68e8209d3c}
}

@misc{SCImago,
  title = {{SCImago} {Journal} \& {Country} {Rank}},
  author = {{SCImago}},
  url = {https://www.scimagojr.com},
  note = {Accessed: 2026-01-12}
}

@article{jain_awareness_2023,
    title = {Awareness of {Racial} and {Ethnic} {Bias} and {Potential} {Solutions} to {Address} {Bias} {With} {Use} of {Health} {Care} {Algorithms}},
    volume = {4},
    issn = {2689-0186},
    url = {https://doi.org/10.1001/jamahealthforum.2023.1197},
    doi = {10.1001/jamahealthforum.2023.1197},
    number = {6},
    urldate = {2026-01-13},
    journal = {JAMA Health Forum},
    author = {Jain, Anjali and Brooks, Jasmin R. and Alford, Cleothia C. and Chang, Christine S. and Mueller, Nora M. and Umscheid, Craig A. and Bierman, Arlene S.},
    month = jun,
    year = {2023},
    pages = {e231197},
}

@article{diao_public_2025,
    title = {Public {Opinion} on {Use} of {Race} in {Clinical} {Algorithms}},
    issn = {2168-6106},
    url = {https://doi.org/10.1001/jamainternmed.2025.6929},
    doi = {10.1001/jamainternmed.2025.6929},
    urldate = {2025-12-26},
    journal = {JAMA Internal Medicine},
    author = {Diao, James A. and Movva, Rajiv and Cheng, Lingwei and Kadoma, Kowe and Shah, Aashna and Powe, Neil R. and Ferryman, Kadija and Manrai, Arjun K. and Pierson, Emma},
    month = dec,
    year = {2025},
}

\newpage
\appendix

\section{Estimating Implicit Contribution of SDoH}
\label{app:phi_estimation}

To quantify $\phi_k^{\mathrm{implicit}}$ for GLI-Global, we exploit the relationship in Equation (4). For each racial/ethnic group $k$, we construct a family of adjusted predictions indexed by $\phi \in [0, 1]$:
\begin{equation}
\widehat{\mathrm{LF}}_{\mathrm{adj}}(X_i; \phi) = \widehat{\mathrm{LF}}^{[\mathrm{GLI\text{-}2012}]}_k(X_i) + \phi \cdot \bigl(\widehat{\mathrm{LF}}^{[\mathrm{GLI\text{-}2012}]}_p(X_i) - \widehat{\mathrm{LF}}^{[\mathrm{GLI\text{-}2012}]}_k(X_i)\bigr).
\end{equation}

For each $\phi$, we compute $z$-scores using these adjusted predictions while retaining the GLI-Global $L$ (skewness) and $S$ (coefficient of variation) parameters:
\begin{equation}
z_{\mathrm{adj}}(X_i, \mathrm{LF}_i; \phi) = \frac{\bigl(\mathrm{LF}_i \,/\, \widehat{\mathrm{LF}}_{\mathrm{adj}}(X_i; \phi)\bigr)^{L} - 1}{L \cdot S}.
\end{equation}

We estimate $\phi_k^{\mathrm{implicit}}$ as the value that minimizes the mean squared difference between adjusted $z$-scores and GLI-Global $z$-scores:
\begin{equation}
\hat{\phi}_k^{\mathrm{implicit}} = \underset{\phi \in [0,1]}{\arg\min} \; \frac{1}{n_k} \sum_{i: E_i = k} \bigl(z_{\mathrm{adj}}(X_i, \mathrm{LF}_i; \phi) - z_{\mathrm{Global}}(X_i, \mathrm{LF}_i)\bigr)^2.
\end{equation}

This calibration procedure recovers $\phi_k^{\mathrm{implicit}}$ because $z$-scores are monotonic transformations of the predicted values, so $z_{\mathrm{adj}}(\cdot; \phi)$ matches $z_{\mathrm{Global}}$ exactly when $\phi = \phi_k^{\mathrm{implicit}}$ as defined in Equation~(4). The MSE is minimized at zero when the adjusted and Global $z$-scores coincide.

We conducted a sensitivity analysis using percent predicted values instead of $z$-scores to assess robustness.

\end{document}